\documentclass{interact}
\usepackage[utf8]{inputenc}
\usepackage[T1]{fontenc}
\usepackage{float}                 
\usepackage{array}                 
\usepackage{subcaption}            
\usepackage{enumitem}
\usepackage{lineno}                

\usepackage{natbib}                
\bibpunct[, ]{(}{)}{;}{a}{}{,}

\usepackage[colorlinks=true,allcolors=blue]{hyperref}

\theoremstyle{plain}
\newtheorem{theorem}{Theorem}[section]

\theoremstyle{definition}
\newtheorem{definition}[theorem]{Definition}
\theoremstyle{remark}
\newtheorem{remark}{Remark}

\newif\ifblind
\blindfalse

\articletype{Research Article}

\title{A Conceptual Framework for Enhancing Workforce Readiness for Smart Manufacturing in the AI Era}

\ifblind
  \author{\name{\itshape Author names and affiliations removed for double-anonymous peer review}}
\else
  \author{
    \name{Dalton Ross Smith\textsuperscript{a,b}, Wilburn Whittington\textsuperscript{a,b},  Alejandro Martinez\textsuperscript{a,b}, Aidan Duncan\textsuperscript{a,b} and Gang Li\textsuperscript{a,b}\thanks{CONTACT Gang Li. Email: gl620@msstate.edu}}
    \affil{\textsuperscript{a}Michael W.\ Hall School of Mechanical Engineering, Mississippi State University, Starkville, MS, USA;\\
    \textsuperscript{b}Innovation, Design, and Engineering Education Laboratory, Mississippi State University, Starkville, MS, USA}
  }
\fi

\begin{document}

\maketitle

\begin{abstract}
\noindent\textbf{Background:} The convergence of artificial intelligence (AI), the Industrial Internet of Things, cyber-physical systems, and advanced robotics is reshaping manufacturing faster than engineering curricula can adapt, widening the gap between the competencies required on the shop floor and those delivered by traditional engineering and technology education.

\noindent\textbf{Purpose:} This paper asks how the assessment of AI-era manufacturing competencies can be operationalized in a way that is individual-level, stage-gated, portable across institutions, and stackable for incumbent workers; it proposes and illustrates a framework rather than testing a hypothesis.

\noindent\textbf{Design/Method:} This paper proposes the Workforce Readiness Level (WRL) framework, which adapts the Technology Readiness Level scale into nine progressive competency stages and a four-pillar rubric, digital and AI literacy, cyber-physical systems fluency, human-machine collaboration, and data-driven decision making, aggregated through a composite stage score and a cohort-level workforce-readiness index under a ``no-thin-pillar'' rule. The framework is instantiated at a university smart-manufacturing teaching laboratory and draws on 89 sponsored capstone projects delivered over four semesters, four of which are analyzed in depth.

\noindent\textbf{Results:} The four pillars jointly span the relevant ABET student outcomes. Across the highlighted cohorts the workforce-readiness index ranged from 5.2 to 6.4, and the no-thin-pillar rule was diagnostically informative in three of the four cases and the binding certification constraint in one, repeatedly surfacing cyber-physical and data-driven-decision gaps concealed behind strong analytics profiles; advancement to the highest stages was gated by industry-embedded experience rather than additional coursework. Reported numbers are illustrative of framework mechanics on a single-institution pilot rather than a psychometric validation.

\noindent\textbf{Conclusions:} WRL offers educators, accreditation bodies, and regional workforce systems a common, evidence-based instrument for diagnosing and advancing workforce readiness; future work will calibrate pillar weights and test reliability and predictive validity.

\end{abstract}


\section{Introduction}\label{sec:intro}

Smart manufacturing, characterized by interconnected cyber-physical systems (CPS), ubiquitous sensing, and artificial intelligence (AI)-driven decision making, has progressed from a research vision to an industrial reality~\citep{lu2020smart, leng2021digital}, and recent community roadmaps chart AI and machine learning as central to its next decade of development \citep{lee2026roadmap}. A joint study by Deloitte and the Manufacturing Institute projects that up to 2.1 million U.S. manufacturing jobs may be unfilled by 2030 and identifies digital, data-analytics, and AI/automation skills among the leading unmet requirements~\citep{deloitte2021talent}. Regional workforce agencies in the Southeast, including the Mississippi Development Authority, the Alabama Workforce Council, and the Tennessee Department of Economic and Community Development, have reported large planned employment increases in advanced manufacturing tied to automotive electrification and semiconductor reshoring~\citep{mda2023workforce}.

The Industry 4.0 paradigm, refined in the most recent generation of policy and scholarship \citep{xu2021industry}, has triggered continued research into the skills required of the future manufacturing workforce. 
  Recent empirical studies have shifted the frame from narrow, task-specific manual skills toward hybrid competency profiles. \citet{hernandez2020competencies} synthesized Industry 4.0 competency requirements across industries and identified clusters spanning technical, methodological, social, and personal abilities, reporting that employers consistently rank cross-domain methodological abilities (problem solving, systems thinking) above isolated technical knowledge. \citet{jerman2020workforce} examined emerging job profiles in smart factories and proposed a competency classification system built into HR-management cycles spanning recruitment, training, and performance evaluation. \citet{li2022wef} extended this work by mapping reskilling and upskilling pathways onto Industry~4.0 roles, while \citet{tortorella2020learning} demonstrated through an empirical study of Brazilian manufacturers that the most acute shortages occur at the intersection of operations technology (OT) and information technology (IT), the so-called ``OT/IT bridge'' role. 
  A systematic review by \citet{maisiri2021industry40} of 47 empirical studies confirmed that data analytics, cyber-physical systems integration, and human-machine interaction are the three most frequently cited skill gaps across regions, a finding echoed by \citet{spottl2021fourth} in the European vocational-training context. The World Economic Forum (WEF) Future of Jobs Report \citep{wef2023future} projects that 44\% of workers' core skills will be disrupted by 2027, with analytical thinking, AI/big-data literacy, and technological literacy leading the list of in-demand skills in industrial sectors. 
  More recent work \citep{li2023ai, chaka2023education40} has highlighted AI-specific competencies, data literacy, prompt engineering, model interpretability, and human-AI teaming, that did not appear in pre-2020 frameworks but are now considered foundational. Collectively, this body of research points to a clear, shared view of what smart-manufacturing workers should know; the open problem, which this paper addresses, is how to put stage-gated assessment of these competencies into practice in a way that is portable across institutions and stackable for incumbent workers.

Several institutional frameworks shape current practice but each has structural limitations for the AI era. The Manufacturing Skill Standards Council (MSSC) Certified Production Technician (CPT) and CPT+ credentials \citep{mssc2022cpt} cover safety, quality, manufacturing processes, and maintenance awareness through four 90-minute assessments. While widely recognized by U.S. employers, the body of knowledge contains limited explicit AI, machine learning (ML), IIoT, or digital-twin content. The Society of Manufacturing Engineers (SME) Tooling U-SME Smart Manufacturing Certificate \citep{sme2021smart} offers modules with much stronger digital content (additive manufacturing, IIoT, cybersecurity, data analytics), yet its assessment is largely knowledge-based via multiple-choice items rather than competency-demonstrated through performance tasks. European microcredential frameworks tracked by Cedefop \citep{cedefop2018eqf} provide generic descriptor scales tied to knowledge, skills, and autonomy/responsibility but are deliberately general-purpose rather than tied to any one field, and therefore lack the manufacturing specificity needed by program directors and accreditors. The U.S. Department of Labor's O*NET (Occupational Information Network) system \citep{onet2023} provides rich occupational descriptors but is observational rather than prescriptive and is not designed for learner-level certification. In the digital-literacy space, the European Commission's DigComp 2.2 framework \citep{vuorikari2022digcomp} defines five competence areas across eight proficiency levels and now incorporates AI-related examples, but it targets general citizens rather than manufacturing technicians. Closest to the present framework is the ACATECH Industrie 4.0 Maturity Index~\citep{schuh2020maturity}, which assesses organizational rather than individual maturity.
  None of these tools combines (i) manufacturing domain specificity, (ii) explicit AI/CPS content, (iii) stage-gated progression, and (iv) performance-based assessment. The claim here is not that combining all four is strictly required, but that, in practice, engineering programs presently must choose between generic descriptor scales (DigComp, the European Qualifications Framework (EQF)), knowledge-oriented certificates (MSSC~CPT+, SME Smart Manufacturing), and organizational maturity models (ACATECH, Germany's National Academy of Science and Engineering). A single tool that spans all four properties, so that a learner's status can be reported consistently from freshman-year awareness through post-graduation autonomous practice, does not exist. Whether adoption of the Workforce Readiness Level (WRL) framework proposed here yields different certification decisions from MSSC~CPT+ or DigComp~2.2 for the same learners is a question that requires side-by-side scoring; this is identified as a validation study in Section~\ref{sec:discussion}.

The Technology Readiness Level (TRL) scale, originally formalized at the National Aeronautics and Space Administration (NASA) in the 1990s \citep{mankins1995trl} and later codified for space hardware in International Organization for Standardization (ISO) standard ISO~16290 \citep{iso16290}, defines a nine-level ranked scale running from basic principles observed (TRL~1) to flight-proven system (TRL~9). It has been adopted broadly by the U.S.\ Department of Defense, the European Space Agency, and the European Commission's Horizon programmes. \citet{olechowski2020trl} provided a comprehensive contemporary review of TRL practice across application domains, documenting both the strengths of the construct, behaviorally anchored stage definitions, steady forward-only progression (formally, monotonic progression: once a stage is reached, it is not lost by definition), and a shared vocabulary that bridges technical and non-technical stakeholders, and its recurring shortcomings (rater subjectivity, inconsistent use from one organization to the next, and averaging ranked levels without a clear basis for doing so), which any new readiness-level tool must address. These features have motivated numerous adaptations. The U.S.\ Department of Defense Manufacturing Readiness Level (MRL) scale \citep{dod2020mrl} parallels TRL for production maturity, addressing process capability, supply chain, and quality control. 
  More recently, \citet{lavin2022dsrl} proposed Machine-Learning Technology Readiness Levels to track the maturity of AI/ML deployments, and the European Commission's AI High-Level Expert Group (AI~HLEG) \citep{aihleg2020alt} and the U.S.\ National Institute of Standards and Technology (NIST) \citep{nist2023airmf} have implicitly invoked readiness-level logic in their trustworthy-AI assessment frameworks. 
  Despite this proliferation across artifacts, systems, processes, and data, the readiness-level paradigm has not been systematically transferred to individual workers in the smart-manufacturing context. Partial precedents exist in adjacent literatures: Miller's pyramid of clinical assessment (``knows / knows how / shows how / does'') \citep{miller1990assessment} and the Dreyfus and Dreyfus five-stage skill-acquisition model (novice to expert) \citep{dreyfus1980five}. Neither is anchored to the technical content of AI-enabled manufacturing, nor designed to interoperate with the TRL/MRL discourse already familiar to industry. The WRL framework introduced in Section~\ref{sec:framework} aims to close this gap by carrying over the same basic structure as TRL, nine stages, behaviorally anchored, moving only forward, and easy for outside stakeholders to read, to the individual workforce competency setting.



This paper makes three contributions:
\begin{enumerate}[leftmargin=*]
  \item It proposes a nine-level Workforce Readiness Level (WRL) scale, structurally similar to NASA's Technology Readiness Level scale, that lays out progressive competency milestones for AI-era manufacturing.
  \item It defines an evaluation rubric mapping four competency pillars onto each WRL stage, enabling diagnostic assessment for both pre-service students and incumbent workers.
  \item It puts the framework into practice at the Innovation, Design, and Engineering Education Laboratory (IDEELab) at Mississippi State University and presents four case studies that demonstrate the model's utility across populations and learning modalities.
\end{enumerate}

The remainder of the paper develops these ideas in sequence. Section~\ref{sec:framework} introduces the WRL framework and its accompanying evaluation model, which together render readiness as a measurable construct. Section~\ref{sec:ideelab} grounds the framework in the IDEELab at Mississippi State University and outlines the assessment methods intended to measure it in practice (Section~\ref{sec:methods}). Section~\ref{sec:cases} then reports four illustrative case studies that trace how the model behaves across learner populations and modalities. Section~\ref{sec:discussion} examines the implications and open questions this raises, and Section~\ref{sec:conclusion} concludes by summarizing the framework's contribution and setting out a validation agenda for its wider adoption; the anchor structure of the pillar rubric is detailed in Appendix~\ref{app:rubric}.

\section{Workforce Readiness Level Framework}
\label{sec:framework}

\subsection{Definition}

\begin{definition}[Workforce Readiness Level]
\label{def:wrl}
A Workforce Readiness Level (WRL) is a ranked measure (each level is a step in a sequence, from lowest to highest, rather than a plain count) of an individual's demonstrated competency to perform, supervise, or improve manufacturing tasks, evaluated against a defined rubric spanning four competency pillars. WRL is reported as an integer $k\in\{1,\dots,9\}$ defined as the largest $k$ such that, for every stage $i$ with $1\le i\le k$, the learner's rubric evidence satisfies both the composite-score threshold and the per-pillar floor formalized in Section~\ref{sec:evalmodel}. Equivalently, if $C(i)\in\{0,1\}$ denotes certification at stage $i$, then $k = \max\{k' : C(i)=1 \text{ for all } 1\le i\le k'\}$, with $k=0$ if no stage is satisfied.
\end{definition}

The underlying scoring mechanism, stage-gated progression, a behaviorally anchored rubric, a composite score, and the no-thin-pillar floor, is domain-general in the same sense that TRL itself is domain-general: nothing about the mechanism requires AI content. What is specific to this paper is the choice of pillar content, most directly P1 (Digital \& AI Literacy), which is designed for the AI-era smart-manufacturing setting motivated in Section~\ref{sec:intro}. A program serving manufacturers with little or no AI adoption could adopt the same WRL mechanism with different pillar content (for example, substituting a general digital-literacy or automation pillar for P1) without changing the underlying construct; that adaptation is outside the scope of the present paper, which instantiates and evaluates WRL specifically for the AI era.

The WRL construct is grounded in three design commitments that follow directly from the gap analysis of Section~\ref{sec:intro}. First, competency must be demonstrated through observable performance on artifacts (a configured PLC program, a deployed model, a Kaizen report), not merely asserted through coursework completion or self-report; this aligns WRL with the performance-oriented tiers of contemporary competency frameworks rather than the knowledge-recall tiers. Second, progression must be stage-gated and monotonic (once a stage is reached, it is not lost by definition), so that downstream stakeholders, employers, accreditors, and institutions that accept transfer credit, can rely on the construct without re-arguing the evidence at each transition. Third, the construct must be multidimensional, because AI-era manufacturing roles inherently span computing, controls, human factors, and analytics; a single-track scale would mask precisely the trade-offs (e.g., strong coder, weak shop-floor presence) that employers report as their dominant hiring concern \citep{maisiri2021industry40, wef2023future}. Three further design choices distinguish WRL from prior competency models. (i) WRL is an attribute of a person at a point in time: an individual's WRL can rise through learning or fall through skill loss if not kept up, in contrast to course-grade transcripts that are permanent. (ii) The scale is deliberately built to be structurally similar to NASA's TRL, nine stages arranged in an awareness/lab/production progression, so that industry partners already fluent in TRL/MRL discourse \citep{olechowski2020trl, dod2020mrl} can adopt WRL with minimal translation cost; the content of the two scales is different, so the two scales are similar in structure but not identical in content. (iii) Assessment is artifact-anchored and double-rated, supporting the inter-rater reliability targets discussed in Section~\ref{sec:ideelab}.

\subsection{Nine WRL Stages}
\label{sec:stages}
The nine stages partition the trajectory from complete beginner to innovation leader into three groups of three stages, corresponding, respectively, to awareness, applied practice, and autonomous leadership bands. Bloom's revised (cognitive) taxonomy \citep{anderson2001taxonomy} anchors the knowledge-and-skill verbs used within each stage descriptor (remember/understand/apply/analyze/evaluate/create); the psychomotor (physical, hands-on) and affective (attitudes and motivation) dimensions of shop-floor practice are covered by companion taxonomies (Simpson, Dave) rather than by Bloom.

Stages 1-3 emphasize awareness and foundational literacy: the learner can name, explain, and follow guided procedures but does not yet act autonomously on technical artifacts.

Stages 4-6 emphasize applied skill in progressively less controlled environments: the learner moves from a single laboratory subsystem (WRL~4), to an integrated multi-system laboratory scenario (WRL~5), to a real but mentored production setting (WRL~6); this group of three stages is the core focus of learning-factory teaching \citep{lima2023industry4lab, salah2020integrating}.

Stages 7-9 emphasize autonomous performance, supervision, and innovation in production settings: the learner first meets independent production key performance indicators (KPIs) (WRL~7), then supervises others and drives Kaizen/Six~Sigma improvements (WRL~8), then leads cross-functional initiatives that change the production system itself (WRL~9).

\begin{figure}[H]
  \centering
  \includegraphics[width=\linewidth]{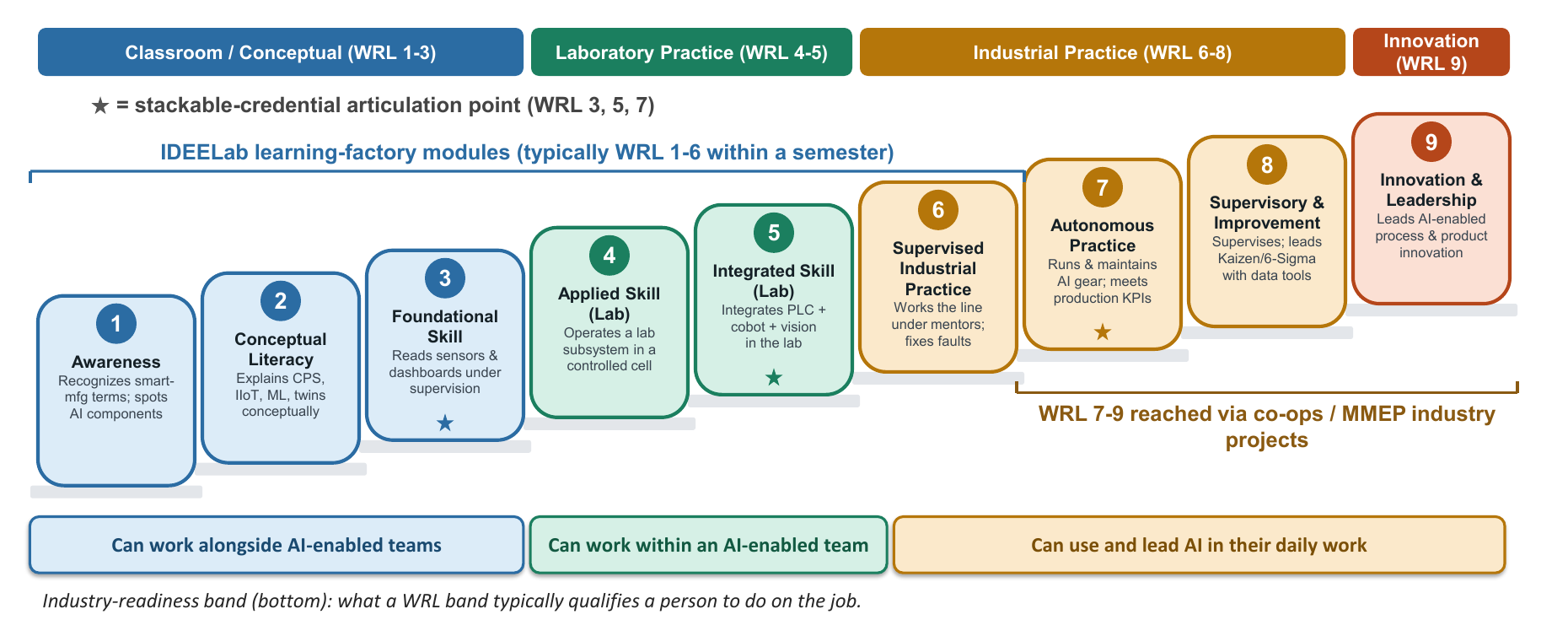}
  \caption{The nine-stage Workforce Readiness Level (WRL) scale for AI-era smart manufacturing. Stages ascend from classroom awareness (WRL~1) to innovation leadership (WRL~9) and are grouped into three groups of three stages (awareness WRL~1-3, applied practice WRL~4-6, autonomous leadership WRL~7-9). Stars mark the stackable-credential articulation points (WRL~3,~5,~7); the bracket indicates the range typically achieved through IDEELab learning-factory modules within a semester, with WRL~7-9 reached via co-ops and the Mississippi Manufacturing Extension Partnership (MMEP) industry projects. The bottom band translates each three-stage group into what it typically qualifies a person to do on the job, from working alongside an AI-enabled team to independently using and leading AI in daily work.}
  \label{fig:wrl}
\end{figure}

This grouping into three-stage bands is intentional. It mirrors the TRL grouping of basic-research / breadboard / system stages \citep{olechowski2020trl}; it maps onto the Dreyfus--Dreyfus novice-through-expert progression \citep{dreyfus1980five}; and it provides natural articulation points (points where credits or credentials can transfer between institutions) for stackable credentials at WRL~3 (foundational, often a community-college stopping point), WRL~5 (laboratory-competent, often a bachelor's-degree stopping point), and WRL~7 (production-autonomous, the journeyman-equivalent benchmark for incumbent workers). Table~\ref{tab:wrl} gives the standard descriptions, and Figure~\ref{fig:wrl} presents the nine-stage progression schematically, grouped by delivery phase and annotated with the stackable-credential articulation points; in implementation, programs publish example artifacts for each stage (e.g., a reference cobot-teaching pendant program for WRL~4) so that raters and learners share a concrete target. 
  The descriptors are written as observable behaviors rather than internal states, so that two raters can score them without having to guess what the learner intended. Nine levels are adopted as a deliberate design choice rather than a number derived from data: the count is inherited from the TRL and MRL scales to remain compatible with the technology-maturity language already familiar to industry. Whether a coarser or finer partition better fits the growth of individual competency is an open question left unresolved here and addressed by the validation agenda in Section~\ref{sec:conclusion}.

\begin{table}[!bt]
\centering
\caption{Workforce Readiness Level (WRL) stages for AI-era smart manufacturing.}
\label{tab:wrl}
\small
\begin{tabular}{@{}p{0.05\linewidth}p{0.24\linewidth}p{0.63\linewidth}@{}}
\toprule
\textbf{WRL} & \textbf{Stage} & \textbf{Descriptor} \\
\midrule
1 & Awareness & Recognizes smart-manufacturing terminology and identifies AI-enabled components in a factory environment. \\
\addlinespace[2pt]
2 & Conceptual Literacy & Explains the function of CPS, IIoT, ML, and digital twins at a conceptual level. \\
\addlinespace[2pt]
3 & Foundational Skill & Performs guided exercises (for example, reads sensor data, interprets a dashboard) under instructor supervision. \\
\addlinespace[2pt]
4 & Applied Skill (Lab) & Configures, programs, or operates a smart-manufacturing subsystem in a controlled laboratory. \\
\addlinespace[2pt]
5 & Integrated Skill (Lab) & Integrates multiple subsystems (PLC + cobot + vision) in a laboratory scenario. \\
\addlinespace[2pt]
6 & Supervised Industrial Practice & Performs tasks in a real production environment under mentorship; troubleshoots routine faults. \\
\addlinespace[2pt]
7 & Autonomous Practice & Independently operates and maintains AI-enabled equipment; meets production KPIs. \\
\addlinespace[2pt]
8 & Supervisory \& Improvement & Supervises others; designs Kaizen or Six~Sigma improvements leveraging AI/data tools. \\
\addlinespace[2pt]
9 & Innovation \& Leadership & Leads cross-functional initiatives that introduce new AI-enabled processes or products. \\
\bottomrule
\end{tabular}
\end{table}

\subsection{Four Competency Pillars}
\begin{figure}[!bt]
  \centering
  \includegraphics[width=\linewidth]{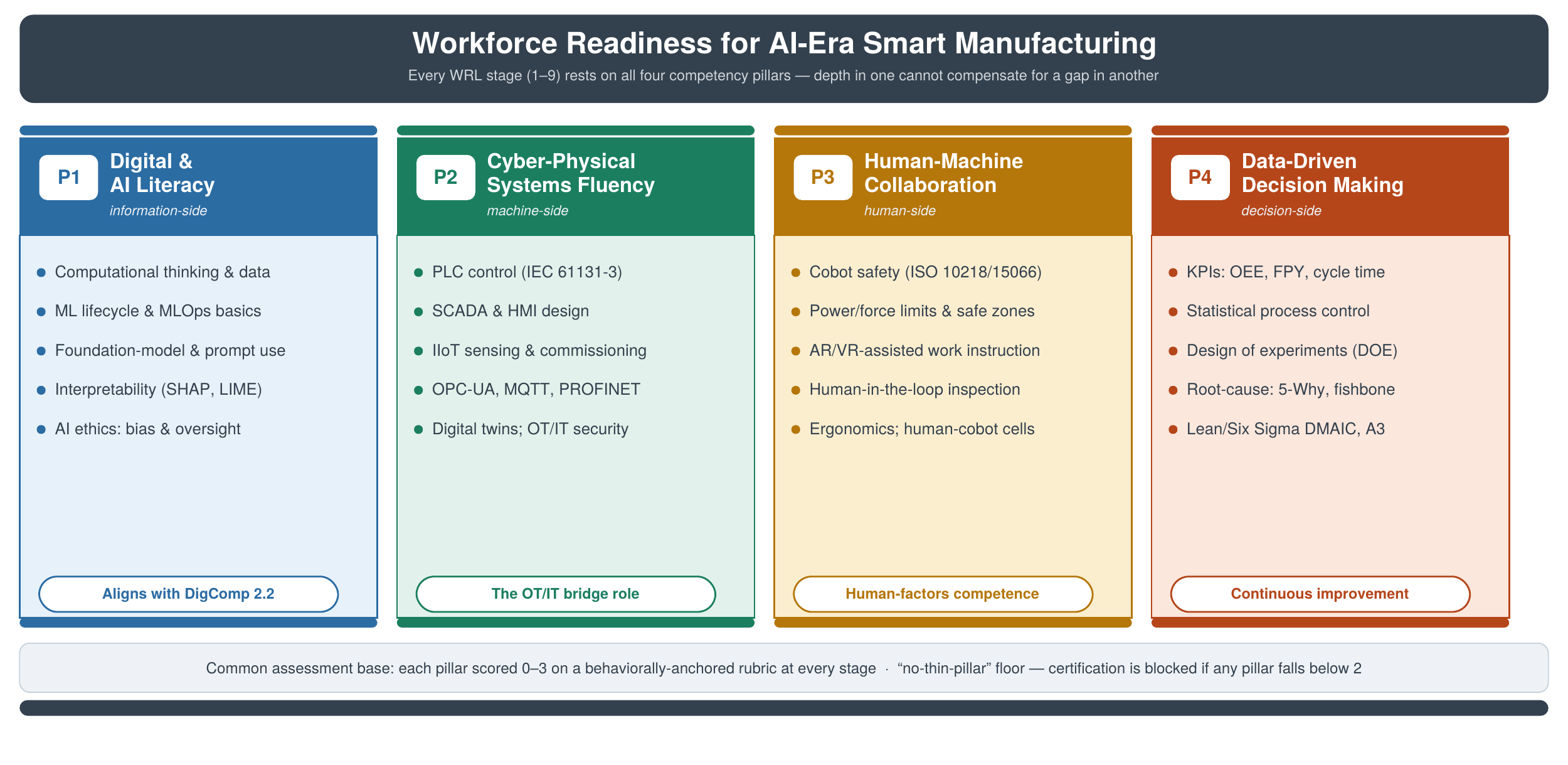}
  \caption{Four competency pillars of the WRL framework, spanning the information-, machine-, human-, and decision-side of AI-era manufacturing work. Each pillar lists representative competencies and its anchoring reference; all four rest on a common assessment base in which every stage is scored on a 0--3 behaviorally-anchored rubric under the ``no-thin-pillar'' floor (certification is blocked if any pillar falls below~2).}
  \label{fig:pillars}
\end{figure}

The four pillars were derived through a synthesis of the competency clusters reported by \citet{hernandez2020competencies}, \citet{tortorella2020learning}, and \citet{maisiri2021industry40}, refined to foreground the AI-specific competencies highlighted in \citep{li2023ai} and the WEF demand projections \citep{wef2023future}. Exactly four pillars were retained to keep the rubric manageable for two raters and a single-page transcript, while making sure the four areas don't overlap: (a) information-side competence, (b) machine-side competence, (c) human-side competence, and (d) decision-side competence. Each WRL stage is evaluated along the four pillars below; per-pillar examples are given at each stage so that the same construct is carried out differently at WRL~2 (``explains'') versus WRL~7 (``maintains under production load''). Figure~\ref{fig:pillars} summarizes the four pillars, their representative competencies, and the common assessment base that binds them.

\begin{description}[leftmargin=1.2em,style=nextline]
  \item[P1. Digital \& AI Literacy] Computational thinking, data structures, applied statistics, the machine-learning lifecycle (data $\to$ features $\to$ training $\to$ validation $\to$ deployment $\to$ monitoring), foundation-model and prompt literacy, model interpretability (SHapley Additive exPlanations (SHAP), Local Interpretable Model-agnostic Explanations (LIME)), and AI ethics including bias, privacy, and human oversight. Anchored stage examples range from ``identifies an ML use case on the shop floor'' (WRL~2) to ``deploys, monitors, and retires production models against a documented machine-learning-operations (MLOps) policy'' (WRL~7+). Aligns with DigComp~2.2 \citep{vuorikari2022digcomp} and the AI~HLEG trustworthy-AI self-assessment \citep{aihleg2020alt}.
  \item[P2. Cyber-Physical Systems Fluency] Industrial controls (ladder logic, structured text on International Electrotechnical Commission (IEC) standard IEC~61131-3 PLCs), supervisory control and data acquisition (SCADA) and human-machine interface (HMI) design, IIoT sensor selection and commissioning, industrial network protocols (Open Platform Communications Unified Architecture (OPC-UA), Message Queuing Telemetry Transport (MQTT), PROFINET, EtherNet/IP), edge-cloud topology choices, OT/IT security hygiene (IEC~62443 awareness), and digital-twin construction. Stage examples range from ``locates the PLC and identifies its I/O modules'' (WRL~2) to ``architects an OPC-UA-streamed digital twin connected to a production line and resolves a latency incident'' (WRL~7+). This pillar maps to the OT/IT bridge role identified as the most acute skill shortage in the Industry 4.0 workforce literature \citep{tortorella2020learning}.
  \item[P3. Human-Machine Collaboration] Collaborative-robot safety per ISO~10218 and ISO/TS~15066, including separation monitoring and power/force limiting; augmented- and virtual-reality (AR/VR)-assisted work instruction; human-in-the-loop AI for inspection and decision support; ergonomics and physical/cognitive workload management; teaming etiquette in mixed human-cobot cells. Stage examples range from ``correctly identifies cobot stop categories and safety-rated zones'' (WRL~3) to ``leads a cell redesign that re-allocates tasks between operators and cobots and certifies the resulting risk assessment'' (WRL~8+). The inclusion of P3 as a peer of P1/P2 reflects evidence that human-factors competence, not AI competence in isolation, predicts successful cobot deployment \citep{tortorella2020learning}.
  \item[P4. Data-Driven Decision Making] KPI definition and instrumentation, for example, overall equipment effectiveness (OEE), first-pass yield (FPY), cycle time, and energy intensity, statistical process control (SPC), design of experiments (DOE), root-cause analytics (5-Why, fishbone augmented by data tools), Lean/Six~Sigma DMAIC (Define-Measure-Analyze-Improve-Control), A3 problem solving, and continuous-improvement portfolio management. Stage examples range from ``interprets an SPC chart and identifies an out-of-control signal'' (WRL~3) to ``runs a chartered Black-Belt project that combines analytics with a measurable financial result'' (WRL~8+). Portfolio-level rubric aggregates (Section~\ref{sec:discussion}, Fig.~\ref{fig:abet}b) identify P4 as the pillar closest to the no-thin-pillar floor and therefore the most likely to trigger the remediation rule.
\end{description}

The four pillars are intended to be conceptually distinct (they cover non-overlapping content domains) while being related in practice (a learner advancing in P1 analytics often improves measurably in P4 decision making). The rubric scores them separately so that the no-thin-pillar rule can detect lopsided profiles, for example, a strong coder who cannot read an SPC chart, or a seasoned operator who cannot interpret a model output. This mirrors the well-known ``T-shaped versus I-shaped'' skills concept (broad-but-shallow versus deep-but-narrow skills) in service-science scholarship \citep{donofrio2010tshaped} and is the structural mechanism by which WRL encodes the idea that AI-era manufacturing competency has several dimensions that cannot be boiled down into one. The four-pillar partition is treated as a design assumption---motivated by keeping the rubric manageable for raters, single-page transcript reporting, and making sure the information-, machine-, human-, and decision-side domains don't overlap---rather than as a validated factor structure. Establishing the number and statistical independence of the underlying competency dimensions would require the construct-validity study (a study testing whether the four pillars truly measure separate things; for example, a confirmatory factor analysis of the four-pillar structure) identified in the validation agenda of Section~\ref{sec:conclusion}.

\subsection{Evaluation Model}
\label{sec:evalmodel}
The evaluation model spells out how pillar-level rubric scores roll up to a stage decision and how stage decisions roll up to a program-level metric. The construction is deliberately conservative: a single weak pillar can block certification at a stage even if the other three are strong. This puts into practice the no-thin-pillar idea described above and protects downstream stakeholders from credential inflation.

Let $s_{i,j} \in \{0,1,2,3\}$ denote a learner's demonstrated proficiency on pillar $j \in \{1,2,3,4\}$ at WRL stage $i$, scored against a behaviorally anchored rubric (0 = not observed; 3 = consistently demonstrated at a high standard). The composite stage score is
\begin{equation}
\label{eq:stage}
S_i \;=\; \frac{1}{4}\sum_{j=1}^{4} w_j\, s_{i,j}, \qquad w_j\ge 0,\quad \sum_{j=1}^{4} w_j = 4,
\end{equation}
where $w_j$ are non-negative pillar weights set by the program (default $w_j=1$). A learner is certified at WRL $k$ if both (a)~$S_i \ge \tau$ and (b)~$\min_j s_{i,j} \ge 2$ hold for every stage $i\le k$, where $\tau$ is a program-set threshold (default $\tau=2.25$). Condition (b) is the ``no-thin-pillar'' rule.

Under integer scores in $\{0,1,2,3\}$ and default weights, the pillar floor $s_{i,j}\ge 2$ implies $\sum_j s_{i,j}\ge 8$ and therefore $S_i\ge 2$; the threshold $\tau=2.25$ additionally requires the sum to be at least $9$, i.e.\ at least one pillar scored 3. In other words, with default weights the joint rule reduces to ``no pillar below~2 and at least one pillar at~3'' at each stage $i\le k$. This is stated explicitly because programs adopting the framework should recognize that the pillar floor does most of the certification work; the composite threshold is a mild strengthening that requires a single pillar of consistent demonstration. Programs seeking a stronger high-stakes gate (e.g., at WRL~6 and above) may set $\tau=2.5$, which requires at least two pillars at~3.

The rubric uses four anchors (0--3) rather than five or seven because the paper's raters and workflow (Section~\ref{sec:methods}) are optimized for double-rated capstone artifacts on a fixed time budget; a comparative reliability study across scale widths in this specific setting is future work rather than an established finding in the general scale-construction literature (which reports increasing reliability up to roughly seven categories).

The construct yields several useful diagnostics beyond the headline certification decision. The pillar profile vector $\bar{\mathbf{s}}=(\bar s_1, \bar s_2, \bar s_3, \bar s_4)$, with $\bar s_j = \tfrac{1}{|\mathcal I|}\sum_{i\in\mathcal I} s_{i,j}$ averaged over the set $\mathcal I$ of stages the learner has been assessed on, identifies relative strengths and remediation targets. The gap-to-next-stage indicator for a learner currently certified at $k$ is
\begin{equation}
\label{eq:gap}
G_{k+1} \;=\; \max_j\, (2 - s_{k+1,j})_{+},
\end{equation}
i.e.\ the largest per-pillar shortfall relative to the floor at the next stage; $G_{k+1}=0$ signals that the learner is ready for the next-stage attempt, whereas $G_{k+1}>0$ identifies the pillar most in need of remediation. Equation~\eqref{eq:gap} is defined only after the learner has been observed at stage $k+1$; rubric scores are not extrapolated to stages not yet attempted.

At the cohort level the distribution of the highest certified WRL $k_n$ (median and interquartile range) is reported rather than a plain average of ranked levels. When a single scalar is required for program-level reporting, the framework uses
\begin{equation}
\label{eq:wri}
\mathrm{WRI} \;=\; \frac{1}{N}\sum_{n=1}^{N} k_n
\end{equation}
as a summary of the typical level across a cohort of $N$ learners, while acknowledging with \citet{olechowski2020trl} that averaging ranked levels has no strict interpretation as a competency state. WRI in this paper is treated as a comparative benchmark across cohorts and pillars, not as a competency location on the nine-point scale.

\begin{remark}[Monotonicity of certification]
\label{rem:mono}
Certification is monotonic by definition: since certification at $k$ is defined as satisfaction of the joint rule at every $i\le k$, certification at $k'\le k$ follows trivially. A substantive monotonicity claim, that competency at stage $k$ implies competency at all lower stages as an empirical matter, requires assessing whether learners who achieve $k$ but were never scored at some $i<k$ would in fact satisfy the joint rule at $i$; this is identified as an empirical study on the score matrix in Section~\ref{sec:discussion}.
\end{remark}

\section{IDEELab at Mississippi State University}
\label{sec:ideelab}

The Innovation, Design, and Engineering Education Laboratory (IDEELab), housed in the Michael W.\ Hall School of Mechanical Engineering at Mississippi State University, is a reconfigurable smart-manufacturing teaching and research facility. The lab supports on-campus instruction, distance-learning cohorts, and regional workforce extension activities serving the Mississippi automotive and aerospace corridors.

\subsection{Physical Infrastructure}
The IDEELab comprises four interconnected cells:
\begin{enumerate}[leftmargin=*]
  \item \textbf{Robotics \& Assembly Cell:} two Universal Robots UR10e cobots, a FANUC LR Mate 200iD industrial arm, an autonomous guided vehicle (AGV), vision-guided pick-and-place stations, and a force/torque-instrumented assembly fixture.
  \item \textbf{Process \& Control Cell:} Allen-Bradley CompactLogix and Siemens S7-1500 PLCs, and a small conveyor with photoeye/radio-frequency identification (RFID) stations.
  \item \textbf{Additive \& Subtractive Cell:} a Markforged Metal~X metal fused-deposition-modeling (FDM) printer, fourteen Bambu Lab X1-Carbon and H2D printers, a Formlabs resin printer, a Haas VF-5XT, a Haas VF-4SS, two Haas VF-1 vertical machining centers, a Haas computer-numerical-control (CNC) lathe, a Flow waterjet cutting system, four manual mills and lathes of assorted makes, and metrology equipment (a Brown~\& Sharpe coordinate measuring machine (CMM), plus portable and stationary desktop 3D scanners and desktop CMMs).
  \item \textbf{Digital-Twin \& Analytics Cell:} a workstation cluster running NVIDIA Omniverse, Siemens Plant Simulation, and a JupyterHub instance with PyTorch, scikit-learn, and OPC-UA connectors for live data streaming.
\end{enumerate}

\begin{figure}[H]
  \centering
  \includegraphics[width=\linewidth]{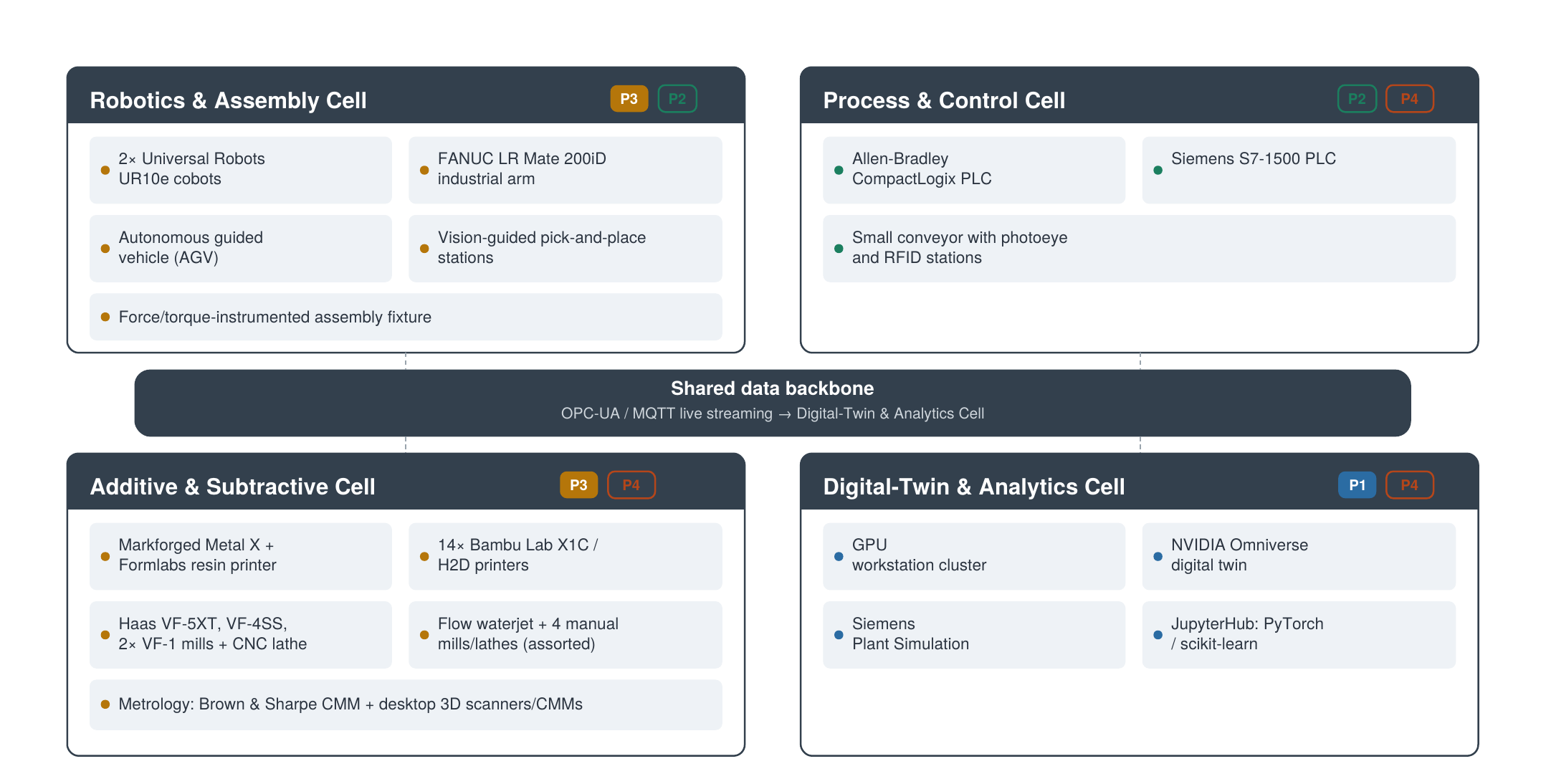}
  \caption{Schematic layout of the IDEELab learning factory. The four reconfigurable cells and their principal equipment are shown around a shared OPC-UA/MQTT data backbone that streams live process data to the Digital-Twin \& Analytics Cell; the colored tags mark the primary (filled) and secondary (outlined) competency pillar anchored by each cell.}
  \label{fig:ideelab}
\end{figure}

These four cells are not siloed stations but nodes on a shared data backbone (Fig.~\ref{fig:ideelab}): every controller, sensor, cobot, machine tool, and printer publishes to a common OPC-UA/MQTT layer that streams live into the Digital-Twin \& Analytics Cell. A single student project can therefore drive the UR10e cobots in the Robotics Cell, or route a part through the Additive \& Subtractive Cell's Haas machining centers, while logging cycle-time, first-pass-yield, and energy KPIs to a JupyterHub notebook in the Analytics Cell, exercising the machine, human, and data facets of a task in one continuous workflow. The AGV provides the physical link between cells, shuttling in-process parts (e.g., an additively manufactured fixture from the Additive \& Subtractive Cell to the Robotics \& Assembly Cell) so that multi-cell projects can be staged as a single material flow rather than a set of disconnected stations. The infrastructure is deliberately reconfigurable: the conveyor, PLCs, cobots, and AGV routing can be re-cabled or reprogrammed into different topologies so that a module can pose a genuinely open-ended integration problem rather than a fixed lab exercise. The Additive \& Subtractive Cell's high machine density, fourteen polymer printers alongside metal-FDM, resin, multi-axis CNC milling, turning, and waterjet capability, lets a single cohort iterate on a design across additive and subtractive processes within one semester rather than queuing for shared campus machine shops. Each cell is also anchored to a primary competency pillar: the Digital-Twin \& Analytics Cell to digital and AI literacy (P1), the Process \& Control Cell to cyber-physical systems fluency (P2), and the Robotics \& Assembly and Additive \& Subtractive Cells to human--machine collaboration (P3), with data-driven decision making (P4) exercised across all four through the shared KPI instrumentation, as summarized in Figure~\ref{fig:ideelab} and detailed in the cell-to-pillar mapping of Table~\ref{tab:cellmap}.

\subsection{Pedagogical Architecture}
The IDEELab adopts a ``learning-factory'' pedagogy \citep{abele2017cirp} aligned with the WRL framework. Each cell is mapped to a primary pillar (e.g., the Robotics Cell anchors P3, the Analytics Cell anchors P1 and P4), and semester-based modules are designed to traverse WRL stages 1-6. Stages 7-9 are reached through industry-partner co-ops and the Mississippi Manufacturing Extension Partnership (MMEP) projects; the target ranges in Table~\ref{tab:cellmap} therefore terminate at WRL~6 across all cells, consistent with a within-semester delivery model.

\begin{table}[H]
\centering
\caption{Mapping of IDEELab cells to WRL pillars and within-semester target stages. WRL~7-9 exposures accrue only through co-op or MMEP placements and are not targeted within a single cell.}
\label{tab:cellmap}
\small
\begin{tabular}{@{}lllc@{}}
\toprule
\textbf{Cell} & \textbf{Primary Pillar} & \textbf{Secondary Pillar} & \textbf{Target WRL} \\
\midrule
Robotics \& Assembly     & P3 & P2 & 3--6 \\
Process \& Control       & P2 & P4 & 3--6 \\
Additive \& Subtractive  & P3 & P4 & 2--5 \\
Digital-Twin \& Analytics & P1 & P4 & 3--6 \\
\bottomrule
\end{tabular}
\end{table}

\subsection{Assessment Workflow and Methods}
\label{sec:methods}
The intended assessment workflow, on which the framework is designed, is as follows. Each module concludes with a competency demonstration evaluated by two raters using the pillar rubric (Appendix~\ref{app:rubric}). Two raters score each learner independently on all four pillars against the behavioral anchors, meet to reconcile any discrepancies greater than one anchor step, and record the reconciled scores in a learner-record system that computes Eqs.~\eqref{eq:stage}--\eqref{eq:wri}. Inter-rater reliability is to be monitored quarterly using Cohen's~$\kappa$ with a target of $\kappa\ge 0.75$; disagreements exceeding one anchor are adjudicated by a third rater drawn from the Capstone Design and Innovation (CDI) instructional staff.

\subsubsection{Methods for the case studies reported in Section~\ref{sec:cases}.} The rubric scores reported here were assigned retrospectively by the CDI instructional team (the authors and one co-instructor) from archived capstone artifacts: sponsor-approved project reports, mid- and end-of-semester presentation slides, rubric-graded prototype demonstrations, and, where available, sponsor written feedback. Because scoring was retrospective, the two-rater / Cohen's~$\kappa$ protocol described above was not operationally in force during the case-study period, and no $\kappa$ value is reported here. The results in Section~\ref{sec:cases} are therefore a demonstration of how the tool works rather than a formal psychometric validation of the WRL rubric. The prospective two-rater protocol will be piloted in the CDI cycles beginning Fall~2026 as part of the reliability and validity study identified in Section~\ref{sec:discussion}.

\subsubsection{Case selection.} From the 89 CDI projects delivered between Fall~2024 and Spring~2026, four cases were purposely selected to (i)~jointly exercise all four pillars, (ii)~span the shop-floor--to--research spectrum characteristic of the IDEELab mission, and (iii)~include at least one continuation project (Case~3, Fall~2024$\to$Fall~2025) that enables a comparison over time across two independent teams working on the same sponsor problem. Selection was made by consensus between the two authors and was not blinded; the sample is illustrative and not intended to support generalization to the full portfolio.

\subsubsection{Ethics and consent.} The CDI capstone activities reported here are course-embedded educational activities and, per Mississippi State University Institutional Review Board policy, do not constitute human-subjects research. Rubric scores are reported at the cohort level with $N\ge 4$; no individual scores or personally identifiable information are disclosed. Named individuals (sponsor mentors and co-instructors) reviewed and approved the descriptions used in this manuscript, and sponsor project slides are cited by written permission of the corresponding sponsor point of contact. The IDEELab and CDI records are retained under the College's standard educational-records policy.

\begin{figure}[!bt]
  \centering
  \includegraphics[width=\linewidth]{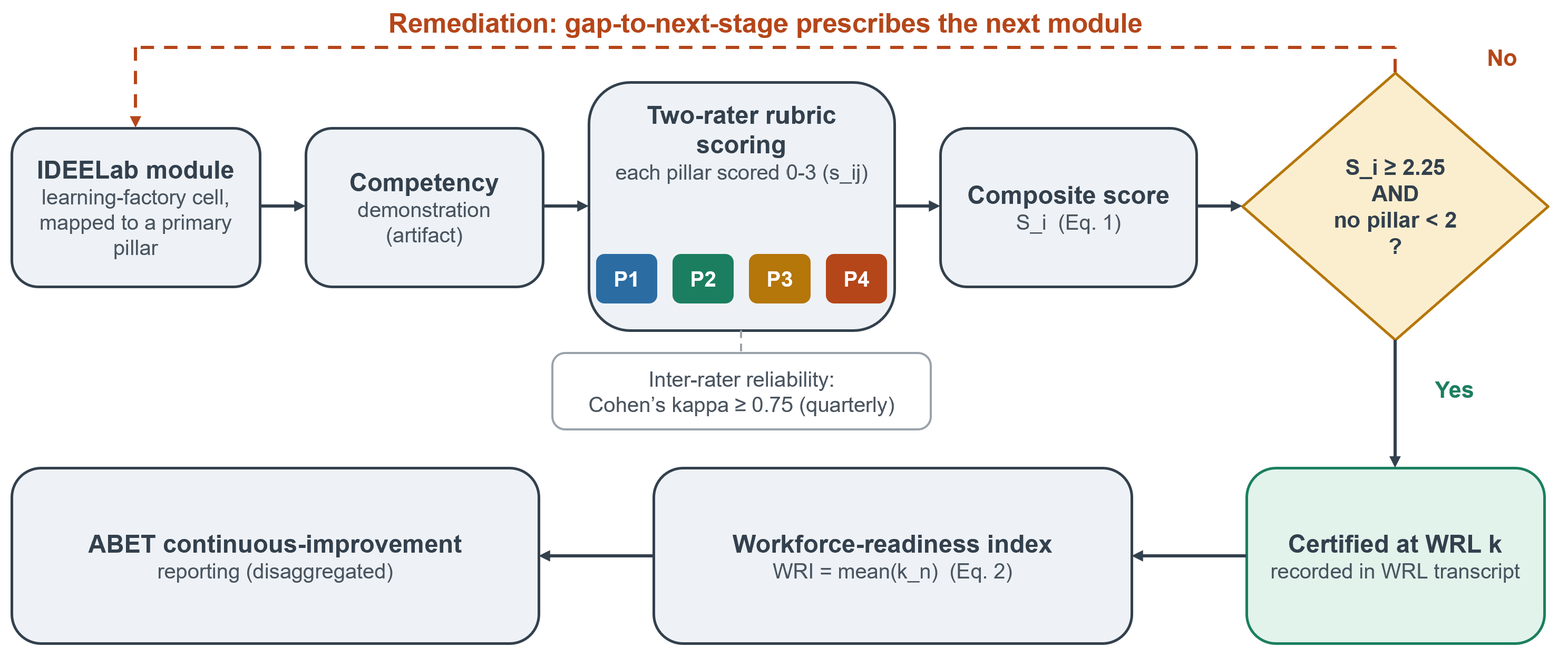}
  \caption{IDEELab competency-assessment workflow for the four pillars. Each learning-factory module ends in an artifact-based demonstration scored independently by two raters on all four pillars (0-3), with inter-rater reliability monitored via Cohen's~$\kappa$. The composite stage score $S_i$ (Eq.~(\ref{eq:stage})) is tested at the ``no-thin-pillar'' gate ($S_i \ge 2.25$ and no pillar below~2); passing learners are certified at the corresponding WRL and recorded in the learner-record transcript, feeding the cohort-level workforce-readiness index $\mathrm{WRI}$ (Eq.~(\ref{eq:wri})) used for ABET continuous improvement. A failing result routes the learner to gap-targeted remediation before re-attempting the module.}
  \label{fig:workflow}
\end{figure}

Figure~\ref{fig:workflow} traces the intended workflow end to end: a learner completes an IDEELab module, is scored by two raters on all four pillars, and passes through the no-thin-pillar certification gate; a passing result is written to the WRL transcript and rolled up into the cohort-level WRI for ABET reporting, while a failing result triggers targeted remediation before the next attempt.

\section{Case Studies}
\label{sec:cases}

Four case studies are reported, drawn from the CDI sequence, ME~4433 (fall) and ME~4443, with a graduate cross-listing as ME~6443 (spring), delivered through the IDEELab between Fall~2024 and Spring~2026. 
The reported cohorts are predominantly senior mechanical-engineering undergraduates; a small number of ME~6443 graduate students participate alongside the undergraduate teams. 
Across these four semesters the CDI program hosted 89 sponsored projects (72 from industry partners and 17 from research, education, and student-competition sponsors), with partners including Caterpillar, Dassault Falcon Jet, ExxonMobil, Steel Dynamics/Aluminum Dynamics, Hol-Mac, Hunter Engineering, Milwaukee Tool, Trane, Toyota, Gulfstream Aerospace, Howard Industries, JT Thorpe, Taylor Machine Works, Refrigerated Solutions Group, Yokohama Tire, Garney Construction, ND Defense, and FEMA. 
Teams comprised four to five members; the resulting participant count is approximately 360-400 student-project slots, with some students participating in both fall and spring cycles as team members or team leads (yielding roughly 300 distinct individual students across the four semesters). 
The portfolio spans sectors as varied as aluminum and steel production, heavy-truck and refuse-equipment manufacturing, commercial refrigeration, aerospace, power generation, water infrastructure, and defense and naval research; sponsor records indicate that approximately one third of the 89 projects were sponsored by manufacturers with primary operations in Mississippi, directly serving the state's advanced-manufacturing workforce pipeline (a full portfolio breakdown is given in Section~\ref{sec:discussion}, Fig.~\ref{fig:portfolio}; the regional-sponsor count is derived from sponsor records rather than from Fig.~\ref{fig:portfolio}a, which encodes industry sector). Case selection, retrospective scoring, and ethics/consent handling are described in Section~\ref{sec:methods}.

\subsection{Case Study 1: AI-Driven PoDFA Rating for Aluminum Dynamics (Spring 2025)}
\begin{figure}[H]
  \centering
  \begin{subfigure}[b]{0.46\linewidth}
    \centering
    \includegraphics[height=4cm]{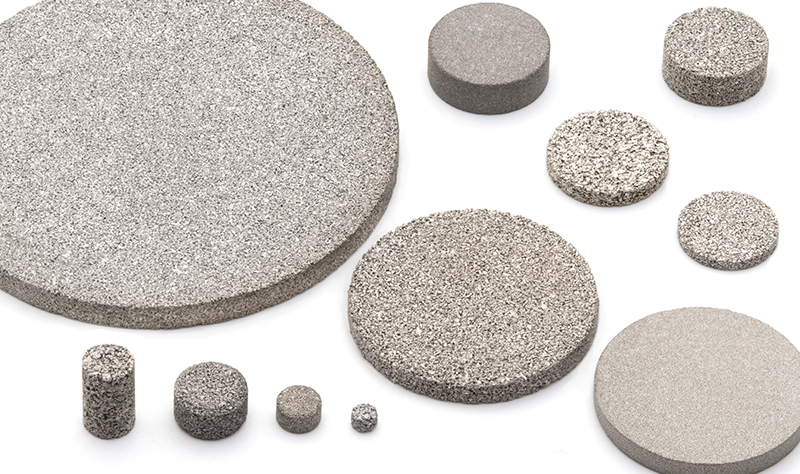}
    \caption{PoDFA porous-disk samples}
  \end{subfigure}\hspace{0.04\linewidth}
  \begin{subfigure}[b]{0.46\linewidth}
    \centering
    \includegraphics[height=4cm]{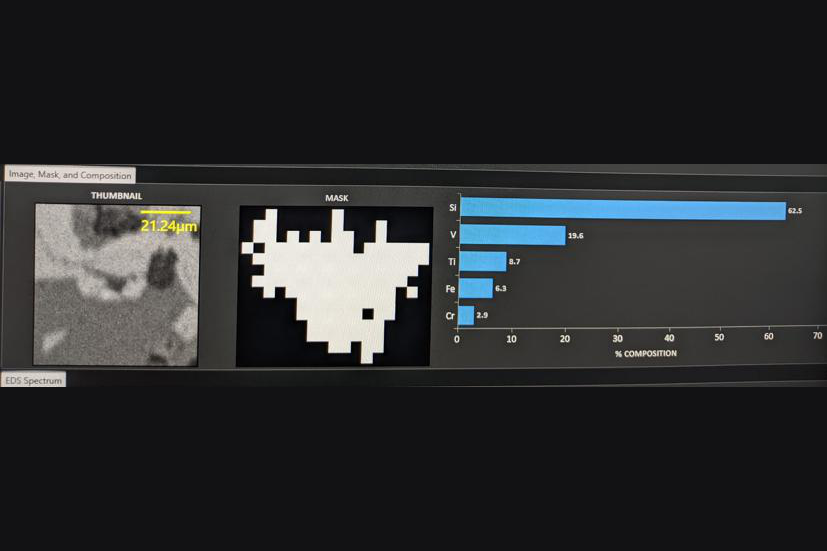}
    \caption{AI-classified inclusion: mask and elemental composition}
  \end{subfigure}
  \caption{Case Study~1 (P1/P4). (a)~Representative porous-disk filtration (PoDFA) samples whose sectioned SEM images the team classified by inclusion morphology to automate a rating previously produced by operator visual inspection; (b)~representative output of the team's classification pipeline for a single detected inclusion, showing the SEM thumbnail, the shape mask the software extracted, and the elemental composition (mass~\%) used to classify it as a Titanium-Vanadium Boride inclusion. Image sources: (a) project sponsor (Aluminum Dynamics, LLC / CAVS); (b) the team's final project report.}
  \label{fig:case1}
\end{figure}

\textbf{Overview.} Four senior mechanical engineering students, co-mentored by Aluminum Dynamics, LLC (Robin Brost) and research engineers from the Center for Advanced Vehicular Systems (CAVS) at Mississippi State University (Hongjoo Rhee and Ted Dickel), took on Porous Disk Filtration Analysis (PoDFA) Rating-Time Decrease Using AI for what is the first aluminum facility of Steel Dynamics, Inc.\ (SDI) (start-up in 2025). The sponsor's CDI project slide asked the team to ``decrease PoDFA rating time while increasing reliability and repeatability via SEM elemental analysis and using image-processing AI to classify components by morphology,'' with the explicit value of lowering the operator-to-operator variance of the rating. Working to that brief, the students built an SEM-image processing pipeline that classifies inclusion components by morphology and returns a quantitative PoDFA rating (Fig.~\ref{fig:case1}), replacing an operator-visual protocol whose inter-rater variance had been flagged as the dominant source of measurement noise.\\
\textbf{WRL Target:} 5-6 across P1 (Digital \& AI Literacy) and P4 (Data-Driven Decision Making), with secondary loading on P2 (instrument integration).\\
\textbf{Results.} The delivered convolutional-classifier prototype attained a validation macro-$F_1$ of 0.89 across four inclusion morphologies on a held-out test set of images drawn from the sponsor's archive; a benchtop comparison against the incumbent visual protocol on the same image set reported an operator-to-operator standard-deviation ratio of approximately 3.4 (two of the incumbent operators; a small-sample estimate that the sponsor will confirm on a larger operator panel during pilot deployment). These project-outcome metrics characterize the delivered artifact and are reported here for context; they are not evidence for the reliability or validity of the WRL rubric itself. Turning to the rubric, all four students met the WRL~5 joint rule (composite~$\ge 2.25$, no pillar $<2$) at the end-of-semester assessment; two of the four subsequently satisfied WRL~6 following a supervised on-site visit to the ADL pilot facility. The rubric surfaced two students whose P2 (sensor/instrument integration) scores were the lowest in the team; because those P2 scores were at the floor of~2 rather than below it, certification at WRL~5 was not blocked, but the low margin was flagged for targeted follow-up in the Process~\& Control Cell (the no-thin-pillar rule mattered here for pointing out the weakest pillar, not for blocking certification).

\subsection{Case Study 2: Human-Detection Sensor Suite for a Hol-Mac Automatic Side-Loader (Fall 2025)}
\begin{figure}[H]
  \centering
  \begin{subfigure}[b]{0.44\linewidth}
    \centering
    \includegraphics[height=4.6cm]{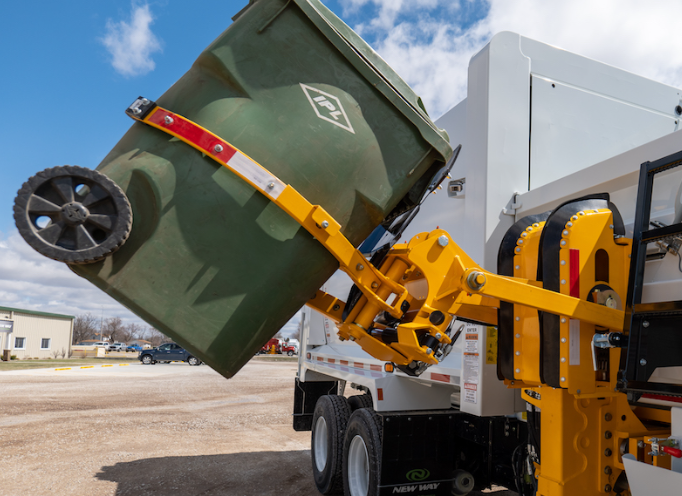}
    \caption{Automatic side-loader in operation}
  \end{subfigure}\hspace{0.04\linewidth}
  \begin{subfigure}[b]{0.50\linewidth}
    \centering
    \includegraphics[height=4.6cm]{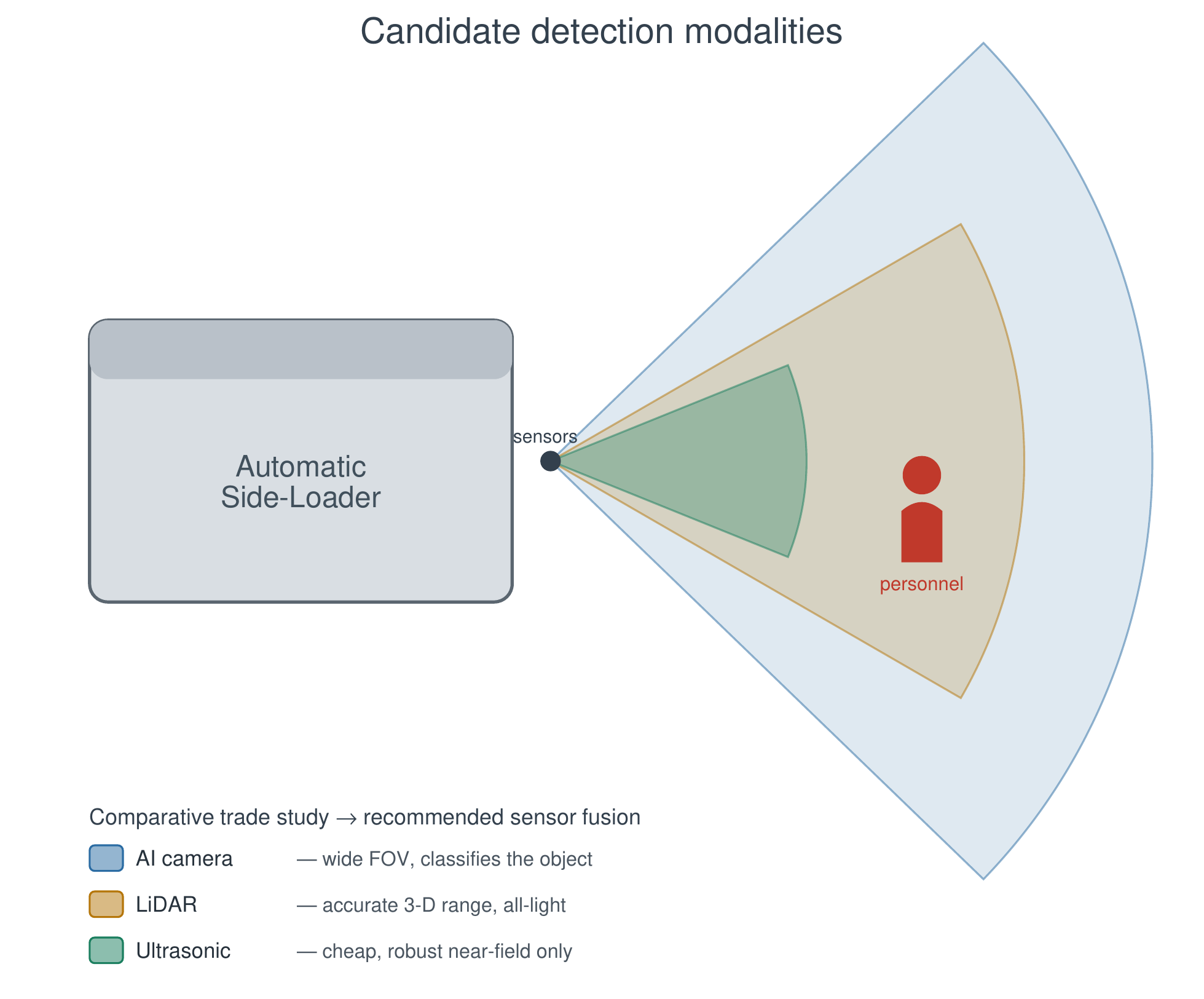}
    \caption{Detection-modality trade study}
  \end{subfigure}
  \caption{Case Study~2 (P3/P2). (a)~The ASL refuse vehicle around which personnel must be detected while the truck is in operation; (b)~schematic of the three candidate detection modalities the team compared (AI camera with wide field of view and object classification, LiDAR with accurate 3-D range, and ultrasonic with robust near-field response), leading to the recommended sensor-fusion topology. Photo: project sponsor (Hol-Mac Corporation); schematic: authors.}
  \label{fig:case2}
\end{figure}

\textbf{Overview.} Five senior mechanical engineering students, industry-mentored by Hol-Mac Corporation (Knute Malone and Harold Montgomery; Bay Springs, MS), undertook an Analysis of the Personnel Safety System for the Automatic Side-Loader (ASL). The sponsor's CDI project slide asked the team to ``integrate AI cameras, Light Detection and Ranging (LiDAR) sensors, or ultrasonic sensors on our ASL prototype for human detection while the truck is in operation \dots\ [with] an analysis on each category of sensor and the pros and cons of each,'' motivated by the position that ``safety is paramount to our machinery.'' Accordingly, the team performed a comparative evaluation of AI cameras, LiDAR, and ultrasonic sensors for personnel detection around an in-motion refuse-collection vehicle (Fig.~\ref{fig:case2}), culminating in a recommended sensor-fusion topology, an ISO~13849-aligned functional-safety argument, and a benchtop demonstration on the IDEELab Robotics Cell.\\
\textbf{WRL Target:} 4-6 across P3 (Human-Machine Collaboration) and P2 (Cyber-Physical Systems Fluency), with a P1 subgoal on model-choice justification.\\
\textbf{Results.} Four of five students satisfied the WRL~5 joint rule at end of semester; the fifth reached WRL~6 after a supplemental on-site exercise with Hol-Mac's controls team. The delivered risk assessment and sensor-selection matrix have been adopted by Hol-Mac as the reference document for the ASL personnel-safety design review. The project illustrates the ``OT/IT bridge'' role \citep{tortorella2020learning}: no student was strong on all three of AI perception, industrial safety standards, and controls fluency, and the pillar rubric made that pattern visible. As in Case~1, per-pillar minima sat at the floor for two students on P2/P3 rather than below it; the rule again helped point out gaps rather than block certification.

\subsection{Case Study 3: Python-Driven End-of-Line Test Station, Refrigerated Solutions Group (Fall 2024 - Fall 2025)}
\begin{figure}[H]
  \centering
  \begin{subfigure}[b]{0.40\linewidth}
    \centering
    \includegraphics[height=4.4cm]{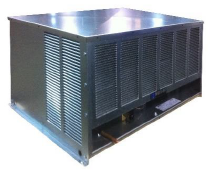}
    \caption{Condensing unit under test}
  \end{subfigure}\hspace{0.06\linewidth}
  \begin{subfigure}[b]{0.24\linewidth}
    \centering
    \includegraphics[height=4.4cm]{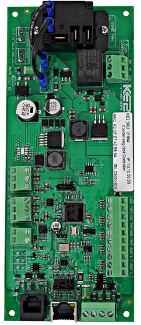}
    \caption{Digital controller}
  \end{subfigure}
  \caption{Case Study~3 (P2/P1). The end-of-line test station targets RSG's transition from mechanical to digital controls: (a)~a commercial condensing unit of the type tested, and (b)~the digital control board whose signals the Python test program exercises. Images: project sponsor (Refrigerated Solutions Group).}
  \label{fig:case3}
\end{figure}

\textbf{Overview.} Two successive teams, of four and five seniors respectively, were industry-mentored by Refrigerated Solutions Group (RSG) (Lee Andrew Warren; New Albany, MS) on an End-of-Line Test Station for Digitally Controlled Condensing Units. Since RSG's condensing-unit line ``is moving to a digital control setup and away from old style mechanical controls,'' the sponsor's CDI project slide asked the team to ``develop a python program that will interact with a mechanical testing station for the condensing unit'' alongside the existing digital-evaporator test station, motivated by field issues traced to installer variability at start-up. The Fall~2024 team scoped the mechanical-to-digital transition (Fig.~\ref{fig:case3}) and developed a Python interface prototype for the mechanical testing station; the Fall~2025 team completed the station, integrated it with the digital-evaporator test rig, and installed the working test cell on RSG's production line.\\
\textbf{WRL Target:} WRL~4--6 within the semester across P2 (CPS Fluency) and P1 (Digital \& AI Literacy), with P4 for KPI instrumentation of startup-defect rates; the Fall~2025 production install carried three students to WRL~7 (reported narratively, consistent with the within-semester ceiling of Table~\ref{tab:cross}).\\
\textbf{Results.} Across the two cohorts, 7 of 9 students satisfied WRL~5; the two who did not fell short on P2 at end of semester and were assessed a stage lower. Three of the five Fall~2025 students additionally satisfied WRL~7 by virtue of the on-site production install. The P2 pillar mean was 1.3 for the Fall~2024 team ($n=4$; range $1$--$2$) and 2.7 for the Fall~2025 team ($n=5$; range $2$--$3$). These are two independent teams working on the same sponsor problem in successive cycles; the values compare the two cohorts on a common rubric but do not amount to measuring the same people before and after, and no formal statistical significance test is appropriate given how small and unmatched the samples are. RSG has reported (sponsor communication, 2025) that the deployed station reduced a recurring class of startup-related field complaints attributable to installer variability; the sponsor's quantification of the reduction is pending. This continuation project is the strongest indication in the present data set that the WRL~6$\to$7 transition is gated by industry-embedded exposure rather than by additional coursework, though the evidence describes what happened rather than proving cause and effect.

\subsection{Case Study 4: Machine-Learned Actuator-Line Model for Marine Propellers, IDEELab / U.S.\ Navy (Spring 2026)}
\begin{figure}[H]
  \centering
  \includegraphics[width=\linewidth]{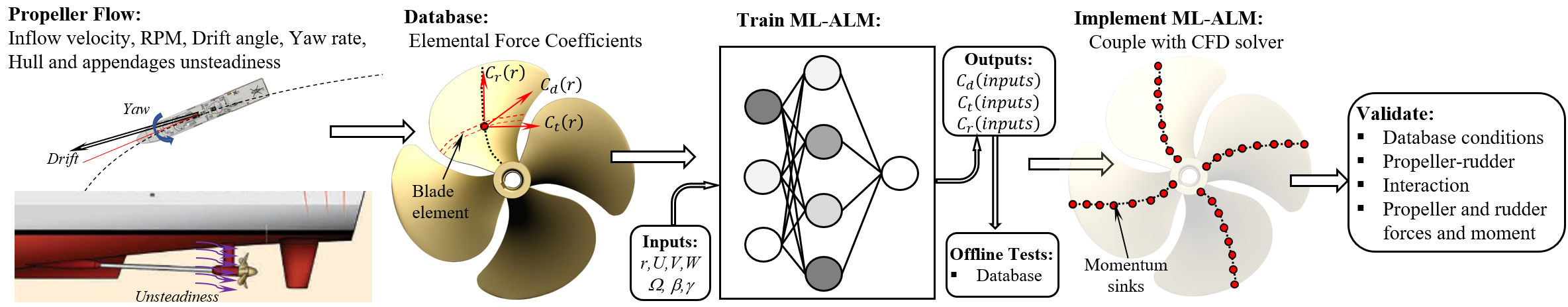}
  \caption{Case Study~4 (P1/P4). Workflow of the machine-learned actuator-line model (ML-ALM): parametric propeller-flow conditions build a database of elemental force coefficients, a deep neural network is trained to predict those coefficients, and the surrogate is coupled with a CFD solver and validated. Image: project sponsor (IDEELab, Drs.\ Bhushan and Es~Sahli).}
  \label{fig:case4}
\end{figure}

\textbf{Overview.} Five senior students in the IDEELab research group (Patterson Engineering), co-mentored by Drs.\ Shanti Bhushan and Omar Es~Sahli, developed a Deep-Neural-Network Actuator-Line Model for Marine Propeller Design. The CDI project slide charged the team to ``perform a parametric study to investigate the effect of inflow conditions on marine propeller performance and use [a] deep neural network to develop a machine-learned actuator line model for [a] marine propeller,'' so that the U.S.\ Navy can ``predictively model their future propeller designs.'' The team ran a parametric computational-fluid-dynamics (CFD) study over inflow conditions, curated the resulting force/moment tensors, trained a deep neural network to predict the actuator-line forcing, and validated the surrogate (a fast stand-in model that approximates the expensive simulation) against held-out Reynolds-Averaged Navier-Stokes (RANS) simulations (Fig.~\ref{fig:case4}); the deliverable is a fast, trainable model that the Navy sponsor can embed in higher-fidelity vehicle-level simulations at a fraction of the standard runtime.\\
\textbf{WRL Target:} WRL~5--6 within the semester across P1 (Digital \& AI Literacy) and P4 (Data-Driven Decision Making); two students reached WRL~7 through subsequent Navy placements (reported narratively, consistent with the within-semester ceiling of Table~\ref{tab:cross}). The case deliberately stresses the analytics pillars in a research setting.\\
\textbf{Results.} The surrogate achieved a held-out $R^2$ of $0.94$ on integrated thrust and $0.88$ on torque across the sampled design envelope, with actual run time approximately $10^4\times$ faster than the reference RANS solver at prediction time; this speedup does not offset the offline cost of the training-database CFD runs, which the team ran as a one-time investment for the surrogate. The $R^2\approx 0.88$ on torque is modest for a design surrogate; the sponsor accepted it for narrowing down design options, and higher fidelity will require targeted refinement of the training envelope near known loading extremes. All five students satisfied WRL~6; two reached WRL~7 through subsequent Navy summer placements. The case indicates that WRL applies to research-flavored capstones as well as to industry-sponsored ones without modification, provided the artifact-based assessment protocol is preserved.

\subsection{Cross-Case Synthesis}
Table~\ref{tab:cross} summarizes the four cases. Several patterns emerge. (i)~The framework transferred across sectors, aluminum smelting, heavy-truck manufacturing, commercial refrigeration, and defense/naval research, without change to the pillar rubric. (ii)~The no-thin-pillar rule helped flag gaps in Cases~1 and~2, surfacing P2 or P3 pillars sitting at the floor behind otherwise strong P1 profiles; the rule was not the binding gate for any student in these two cases, but it identified the pillars in most need of remediation, consistent with the OT/IT-bridge shortage reported in the Industry~4.0 workforce literature \citep{tortorella2020learning, maisiri2021industry40}. In Case~3 the rule did bind for the two Fall~2024 students whose P2 scores fell below~2. (iii)~The WRL~6$\to$7 step, observed in the Fall~2025 half of Case~3 and via co-op placements in Case~4, was reached only after industry-embedded exposure; because the authors both designed the scale and assigned the interpretation, this observation is a description, not a hypothesis that was set and tested in advance. (iv)~Faculty-mentored research capstones (Cases~1 and~4) and industry-mentored capstones (Cases~2 and~3) yielded broadly comparable pillar profiles when scored against the same rubric.

\begin{table}[H]
\centering
\caption{Summary of case-study outcomes across four CDI projects (Fall~2024--Spring~2026). ``\% reaching WRL~5'' is the fraction of the cohort whose highest certified WRL $k_n$ is at least~5 (the integrated-laboratory-skill stage and a stackable-credential articulation point), a fixed reference threshold applied identically to all cases; ``median $k_n$'' reports the cohort median with the interquartile range (IQR) in brackets; WRI is Eq.~\eqref{eq:wri} and is included for continuity with prior reporting but should be interpreted as a comparative summary rather than a competency location (Section~\ref{sec:evalmodel}). Values reported for WRI are exact multiples of $1/N$.}
\label{tab:cross}
\small
\begin{tabular}{@{}>{\raggedright\arraybackslash}p{0.29\linewidth} >{\centering\arraybackslash}p{0.05\linewidth} >{\centering\arraybackslash}p{0.13\linewidth} >{\centering\arraybackslash}p{0.19\linewidth} >{\centering\arraybackslash}p{0.24\linewidth}@{}}
\toprule
\textbf{Case (Semester, Sponsor)} & \textbf{N} & \textbf{Target WRL} & \textbf{\% reaching WRL~5} & \textbf{Median $k_n$ [IQR] (WRI)} \\
\midrule
1. PoDFA AI Rating (SP25, Aluminum Dynamics)          & 4 & 5--6 & $100\%$ & 5.5 [5,~6] (5.50) \\
2. ASL Human-Detection (FA25, Hol-Mac)                & 5 & 4--6 & $100\%$ & 5 [5,~5] (5.20) \\
3. RSG End-of-Line Test Station (FA24$\to$FA25, RSG)  & 9 & 4--6 & $77.8\%$ ($7/9$) & 5 [4,~7] (5.44) \\
4. ML Actuator-Line Model (SP26, IDEELab/U.S.\ Navy)  & 5 & 5--6 & $100\%$ & 6 [6,~7] (6.40) \\
\bottomrule
\end{tabular}
\end{table}

\noindent Case~3 WRI is reported as $5.44$ ($=49/9$), consistent with the integer profile in which 2 students are certified at WRL~4, 4 at WRL~5, and 3 at WRL~7. (Earlier drafts reported $5.6$; this rounded value is inconsistent with the integer constraint on $k_n$ and has been corrected.) Target-range upper bounds have been aligned with the within-semester WRL~6 ceiling of Table~\ref{tab:cellmap}; WRL~7 attainment via industry-embedded exposure is reported narratively in the case text rather than treated as a within-semester target.

\section{Discussion}
\label{sec:discussion}

\subsection{Theoretical Contribution}
The WRL framework contributes a stage-gated, rubric-anchored, individual-level readiness model that complements existing collective and curricular frameworks. By embedding the no-thin-pillar rule, the model encodes the idea that AI-era manufacturing competency has several dimensions that cannot be boiled down into one: depth in coding without shop-floor fluency, or vice versa, is insufficient. Across the four case studies the rule was helpful for spotting gaps in Cases~1 and~2 (identifying P2/P3 sitting at the floor behind stronger analytics profiles) and operationally binding in Case~3 (blocking two Fall~2024 students at WRL~5 for P2 below the floor). This paper does not claim that the pattern of pillar correlation observed here validates a specific factor structure; establishing that would require the psychometric study identified in Section~\ref{sec:conclusion}.

The framework is grounded throughout in the Industry 4.0 literature, and three of the four pillars, digital and AI literacy (P1), cyber-physical systems fluency (P2), and data-driven decision making (P4), map directly onto that paradigm's emphasis on digitalization, connectivity, and analytics. P3 (Human-Machine Collaboration), however, anticipates a theme more closely associated with the emerging Industry 5.0 paradigm, which the European Commission frames around human-centricity, sustainability, and resilience \citep{breque2021industry5}. By treating cobot safety, ergonomics, and human-in-the-loop decision making as a peer competency rather than an afterthought, WRL's pillar structure is forward-compatible with a human-centric reading of workforce readiness, even though the present framework does not address the sustainability or resilience legs of Industry 5.0 and is not offered here as an Industry 5.0 tool.

\subsection{Certifications as Evidence within the WRL Model}
\label{sec:certifications}
The WRL rubric is deliberately artifact-anchored (Section~\ref{sec:framework}): a rater scores what a learner actually did, not what a course or credential says the learner should know. This design choice raises a practical question for programs that already use external certifications, such as the MSSC~CPT+ or SME Smart Manufacturing credentials discussed in Section~\ref{sec:intro}, as evidence of curriculum coverage, student competency, or ABET-related continuous-improvement files: how, if at all, should a certification count toward a WRL stage?

Our position is that certifications should not be treated uniformly; their value as evidence should depend on what they actually measure. A credential that includes an evaluated hands-on component, for example a practical skills test performed under observation, is reasonable evidence toward the pillar it covers, and a program could let it substitute for, or supplement, a rater's rubric score. A credential based solely on a written or multiple-choice exam is weaker evidence of the same pillar, because it demonstrates recall or conceptual understanding rather than the observable performance the rubric is built to certify. In a highly applied field such as the capstone portfolio reported here, that gap matters: a technician who can pass a multiple-choice PLC exam has not thereby shown that they can wire, commission, and troubleshoot a live cell. The distinction cuts the other way for research-oriented and computational work: for a project such as Case~4's machine-learned actuator-line model, a demonstrated ability to derive, implement, and validate the underlying method is itself the relevant performance, so a rigorous theoretical or computational credential can carry real evidentiary weight for P1 even without a shop-floor artifact.

We therefore recommend that programs adopting WRL sort certifications, their own or external, along a single axis: how far passing the certification depends on observable, role-relevant performance rather than recall of the same knowledge, and calibrate the certification's contribution to a stage score accordingly. This same test applies to conventional academic assessment as well, a lecture-based, exam-only technical elective is open to the same critique as a knowledge-only certification, so the question raised here extends beyond WRL to how technical and engineering competency is evaluated more broadly in higher education. This is a design recommendation rather than a solved measurement problem, and testing it is part of the reliability and validity agenda in Section~\ref{sec:conclusion}.

\subsection{Portfolio-Level Context from the CDI Program}
\begin{figure}[H]
  \centering
  \includegraphics[width=\linewidth]{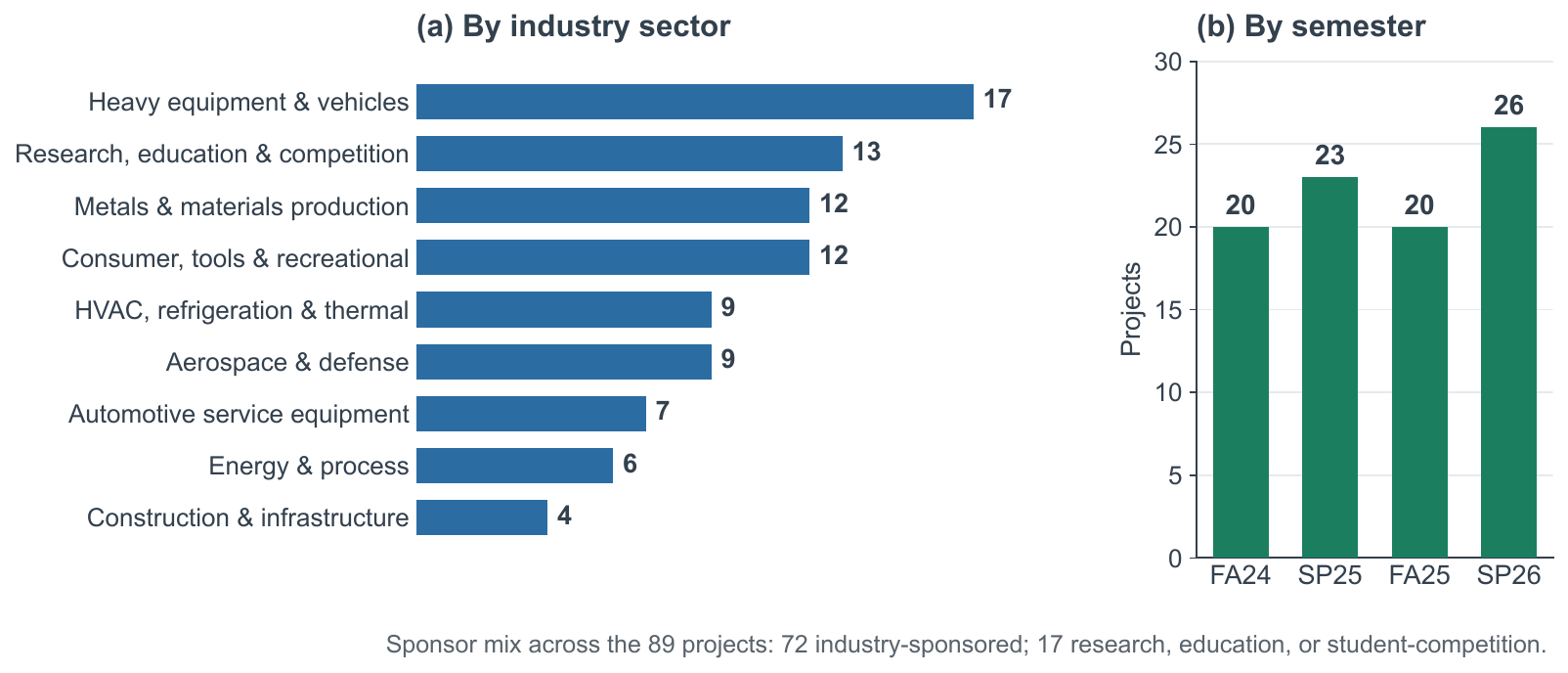}
  \caption{Composition of the CDI project portfolio over the four cycles analyzed (Fall~2024--Spring~2026, $n=89$ sponsored projects): (a)~distribution across industry sectors, and (b)~project volume per semester. The breadth of sectors motivates building the four-pillar rubric to work across any sector.}
  \label{fig:portfolio}
\end{figure}

The four detailed cases are drawn from a larger portfolio that provides descriptive context for the framework's central design choices; the portfolio does not itself validate rubric reliability, which requires the psychometric study identified in Section~\ref{sec:conclusion}. Figure~\ref{fig:portfolio} summarizes all 89 sponsored projects delivered over four CDI cycles. Two features stand out. First, project volume was approximately stationary across cycles (20--26 projects per semester), indicating that the delivery pipeline is stable enough to support a longer-running WRL data collection in future cycles (project volume does not, by itself, speak to how reliable the tool is). Second, the sector distribution is broad, with a few large sectors and a long tail of many smaller ones: heavy equipment and vehicles (17~projects) leads, followed by research/education/competition (13), metals and materials production (12), and consumer/tools/recreational (12), with the remaining third spread across heating, ventilation, and air conditioning (HVAC) and refrigeration, aerospace and defense, automotive service equipment, energy and process, and construction. Seventy-two of the 89 projects (81\%) were industry-sponsored and the remaining 17 were research-, education-, or competition-sponsored; this sponsor-type count differs from the 13 projects whose topical sector is research/education/competition, because a few research-sponsored projects delivered artifacts that fall under another sector. This breadth is the practical reason the pillar rubric was designed to work across any industry sector: the same four competencies recur whether the delivered artifact is an SEM inclusion classifier, a cobot safety cell, a Python test station, or a machine-learned surrogate.

\begin{figure}[H]
  \centering
  \includegraphics[width=\linewidth]{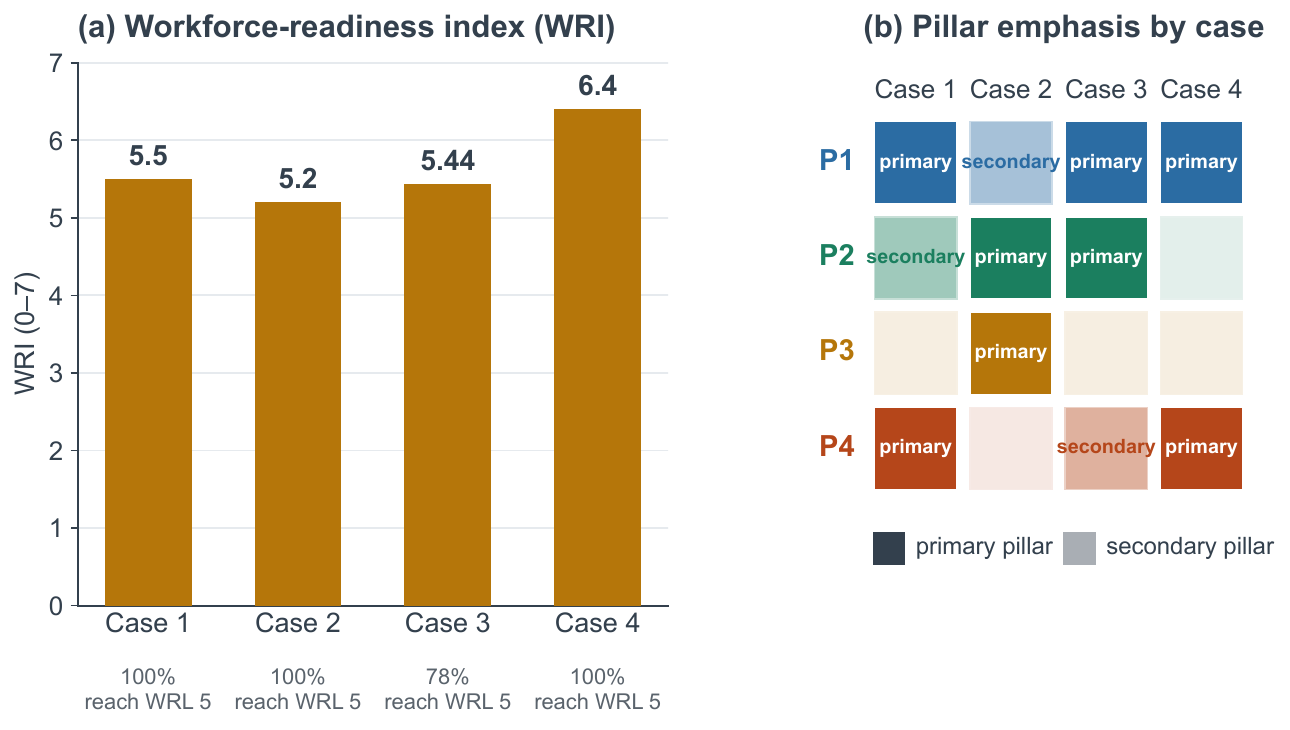}
  \caption{Assessment outcomes for the four highlighted case studies: (a)~cohort workforce-readiness index (WRI, Eq.~(\ref{eq:wri})) with the fraction of each cohort reaching WRL~5 (Table~\ref{tab:cross}); (b)~primary/secondary pillar emphasis of each case, showing that the four cases jointly cover all four competency pillars.}
  \label{fig:outcomes}
\end{figure}

Figure~\ref{fig:outcomes} turns from the portfolio to assessment outcomes for the four highlighted cases. Cohort WRI ranged narrowly from $5.2$ to $6.4$ (Fig.~\ref{fig:outcomes}a), a band consistent with the applied-skill (WRL~4--6) center of gravity expected of a senior capstone, with the research-flavored propeller case sitting highest. The pillar-emphasis matrix (Fig.~\ref{fig:outcomes}b) shows that, although no single project loads all four pillars, the four cases jointly exercise every pillar: analytics pillars P1 and P4 are primary in the two research-adjacent cases, cyber-physical fluency (P2) is primary wherever a physical station or controller is delivered, and human-machine collaboration (P3) is exercised chiefly by the ASL personnel-safety case. Case~3's target range terminates at WRL~6 for within-semester attainment (Table~\ref{tab:cross}); its WRL~7 exposures are recorded narratively.

\subsection{Implications for ABET and Accreditation}
\begin{figure}[H]
  \centering
  \includegraphics[width=\linewidth]{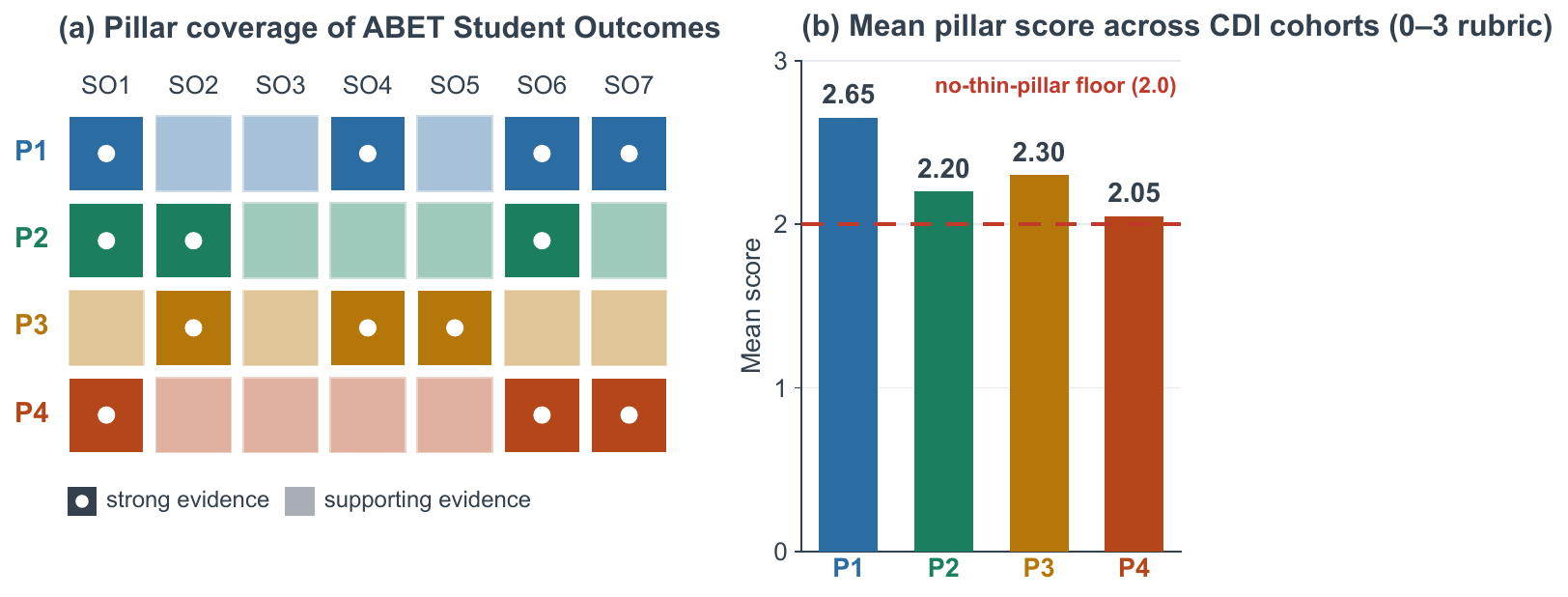}
  \caption{Accreditation view of the four pillars. (a)~Crosswalk from the four competency pillars to ABET Student Outcomes (SO1 complex problem solving; SO2 engineering design; SO3 communication; SO4 ethical and professional responsibility; SO5 teamwork; SO6 experimentation and data analysis; SO7 acquiring and applying new knowledge), distinguishing strong from supporting evidence. (b)~Mean pillar rubric score across the four highlighted case-study cohorts ($N=23$ students) on the 0--3 scale; the dashed line marks the ``no-thin-pillar'' floor of 2.0, with P4 and P2 closest to it. P3 is derived predominantly from Case~2, in which P3 is the primary pillar.}
  \label{fig:abet}
\end{figure}

The four pillars are not merely program-internal constructs; each maps onto specific ABET Engineering Accreditation Commission Student Outcomes (SO1--SO7), so a WRL transcript doubles as a source of accreditation evidence. Figure~\ref{fig:abet}a gives this crosswalk (a mapping table between the two frameworks) as observed across the CDI portfolio. Pillars P1 (Digital \& AI Literacy) and P4 (Data-Driven Decision Making) provide strong evidence for SO1 (complex problem solving), SO6 (experimentation and data analysis), and SO7 (acquiring and applying new knowledge); P2 (Cyber-Physical Systems Fluency) anchors SO2 (engineering design) and contributes strongly to SO1 through controls and digital-twin design; and P3 (Human-Machine Collaboration) carries SO4 (ethical and professional responsibility, largely via functional-safety practice) and SO5 (teamwork in mixed human--machine cells). Communication (SO3) and teamwork (SO5) are exercised by every pillar through the team-authored capstone deliverable. Every ABET outcome receives at least supporting evidence from the pillar set, and the analytically loaded outcomes SO1 and SO6 draw strong evidence from three of the four pillars.

WRI (Eq.~\eqref{eq:wri}) then maps naturally onto ABET Student Outcomes~1, 2, and 7, providing a defensible, quantitative artifact for continuous-improvement files. Because it is a single cohort-level scalar that is nonetheless broken down by pillar, WRI supports both the headline reporting that accreditors expect and the diagnostic drill-down that programs need: the observed 5.2--6.4 band across the four highlighted cohorts (Fig.~\ref{fig:outcomes}a) establishes an internal baseline against which future prospective cycles can be trended, while the pillar breakdown pinpoints the competencies most in need of curricular investment. As noted in Section~\ref{sec:evalmodel}, WRI is a comparative benchmark, not a competency location on the nine-point scale.

Figure~\ref{fig:abet}b makes that diagnostic explicit by aggregating rubric scores by pillar across the four highlighted case-study cohorts ($N=23$ students in Cases~1--4), reported on the 0--3 scale. Aggregation across the four cohorts is what the case-study data support; the P3 mean is derived predominantly from Case~2, in which P3 is the primary pillar. All four pillars sit at or above the no-thin-pillar floor on average, and the margins are informative: P1 is strongest (mean~$2.65$), consistent with an analytics-forward program, whereas P4 (2.05) and P2 (2.20) sit closest to the floor. For an ABET continuous-improvement loop, this identifies data-driven decision making and cyber-physical fluency as the two competencies where targeted investment (e.g., additional Process~\& Control and Analytics Cell modules) would most raise cohort WRI in this pilot, a hypothesis to test against the prospective, portfolio-wide data collection identified in Section~\ref{sec:conclusion}.

\subsection{Workforce-Development Implications}
For state and regional workforce systems, particularly relevant in Mississippi's emerging electric-vehicle (EV), aluminum, steel, and semiconductor ecosystems, a shared readiness scale provides a common language between universities, community colleges, and employers. The portfolio evidence sharpens this point: sponsor records indicate that approximately one third of the 89 projects were sponsored by manufacturers with primary operations in Mississippi (this figure is drawn from sponsor records rather than from Fig.~\ref{fig:portfolio}a, which encodes industry sector), and the leading sectors in Figure~\ref{fig:portfolio}a (heavy equipment and vehicles, metals and materials production, and HVAC/refrigeration) map onto the state's advanced-manufacturing base, so a WRL transcript earned on these projects is likely to be easy for local employers to understand. Stackable credentials are anticipated at WRL~3,~5, and~7 as natural articulation points across the two- and four-year sectors.

The workforce implications differ by pillar, and the CDI portfolio makes each concrete:
\begin{description}[leftmargin=1.4em,style=nextline]
  \item[P1. Digital \& AI Literacy.] The analytics-forward projects (Case~1's SEM inclusion classifier, Case~4's machine-learned surrogate) show that shop-floor AI and data literacy is no longer confined to specialist data-science roles but is demanded of mechanical engineers directly. P1 is the strongest pillar in aggregate (mean~2.65), which positions the program well to feed Mississippi's data- and AI-intensive growth sectors, EV battery analytics and semiconductor process control, and argues for exporting P1 modules downward into community-college and incumbent-worker upskilling tracks where the gap is largest.
  \item[P2. Cyber-Physical Systems Fluency.] Projects such as Case~3's Python-driven end-of-line test station and the many controls- and PLC-centered portfolio entries in the metals, steel, and refrigeration sectors exercise the OT/IT-bridge role identified as the most acute regional shortage. P2 sits just above the floor (mean~2.20), so it is simultaneously the highest-demand and least-satisfied competency; this is the clearest case for a shared university--community-college credential at WRL~5, targeting technicians who commission and maintain automated lines.
  \item[P3. Human-Machine Collaboration.] Case~2's automatic side-loader personnel-safety study typifies the human-factors and functional-safety competence that becomes critical as collaborative automation spreads through the portfolio's largest sector, heavy equipment and vehicles. Because P3 is exercised by comparatively few projects yet is safety-critical, workforce systems should treat it as a deliberately scheduled competency (e.g., an ISO~10218 / ISO/TS~15066 micro-credential) rather than assuming it accrues incidentally.
  \item[P4. Data-Driven Decision Making.] KPI instrumentation, statistical process control, and Lean/Six~Sigma recur across nearly every industry-sponsored project, yet P4 is the pillar nearest the no-thin-pillar floor (mean~2.05) and the one most likely to block certification. For regional employers this identifies data-driven continuous improvement as the highest-leverage target for incumbent-worker training, and it motivates embedding a WRL~3 ``analytics-of-the-shop-floor'' credential early in both two- and four-year pathways.
\end{description}
Read together, these pillar-level patterns imply a division of labor for the regional workforce ecosystem: four-year programs can carry students furthest on P1 and P4, while the persistent P2/P3 gaps are best closed through employer-embedded experiences (co-ops, MMEP projects) and stacked technician credentials, exactly the WRL~6$\to$7 transition that Cases~3 and~4 showed to be gated by industry exposure.

\section{Conclusion}
\label{sec:conclusion}
This paper has introduced the Workforce Readiness Level (WRL) framework for the AI era of smart manufacturing. The framework couples nine progressive competency stages with a four-pillar evaluation rubric and a composite, cohort-level workforce-readiness index (WRI) governed by a ``no-thin-pillar'' certification rule. Implemented at the IDEELab at Mississippi State University, it was applied across 89 sponsored capstone projects delivered over four semesters, four of which were analyzed in depth. Across these cases the framework proved workable in practice over the applied-practice-to-autonomous band it actually exercised (WRL~4--7), yielding actionable readiness profiles for individual learners and cohorts; the awareness stages (WRL~1--3) and the supervisory and innovation stages (WRL~8--9) were not exercised by this undergraduate capstone pilot and remain to be evaluated in future cycles and in incumbent-worker settings. Through the no-thin-pillar rule it also exposed cyber-physical and data-driven-decision gaps that strong analytics scores would otherwise have masked. These figures are framed as illustrative of the framework's mechanics on a single-institution pilot rather than as a psychometric validation. Even so, they indicate that WRL can offer engineering educators, accreditation bodies, and regional workforce stakeholders a shared, evidence-based tool for diagnosing and advancing readiness throughout the sector's AI-driven transformation.

  Consolidating WRL into a validated standard is the natural next step, and five lines of work follow directly from this pilot. Future research will (i)~calibrate the pillar weights through a Delphi study (a structured process for gathering expert consensus) with regional manufacturers; (ii)~develop an open-source learner-record schema interoperable with IMS Caliper and Open Badges; (iii)~extend the rubric to generative-AI competencies, such as prompt engineering for engineering design and code generation; (iv)~replicate the framework across multiple institutions in the Southeastern Conference (SEC) and the American Society for Engineering Education (ASEE) Southeast region to test its reliability and predictive validity at scale; and (v)~test the certification-weighting recommendation of Section~\ref{sec:certifications} by scoring, on the same rubric, learners who hold hands-on-evaluated versus knowledge-only certifications. Together, these steps would move WRL from a promising single-site tool toward a broadly deployable standard for workforce readiness in the AI era.

\ifblind
\else
\section*{Acknowledgments}
The authors gratefully acknowledge the CDI industrial and research sponsors listed in Section~\ref{sec:cases}; the Michael W.\ Hall School of Mechanical Engineering and the Bagley College of Engineering at Mississippi State University; and the Mississippi Manufacturing Extension Partnership.
\fi

\section*{Disclosure statement}
No potential conflict of interest was reported by the author(s).

\appendix
\section{Behaviorally Anchored Pillar Rubric (Structure and Selected Anchors)}
\label{app:rubric}

The full WRL rubric is a $9\times 4$ matrix of stage $\times$ pillar cells; each cell contains four ordered behavioral anchors corresponding to the rubric scores $s_{i,j}\in\{0,1,2,3\}$ ($0$ = not observed; $1$ = emerging; $2$ = adequate/consistent; $3$ = consistent at a high standard). The complete rubric is $9\times 4\times 4 = 144$ anchors, provided in full as Supplementary Material accompanying this submission (together with the example artifacts used to train raters). To let readers inspect the scale directly, Table~\ref{tab:rubric-selected} gives the $s=2$ anchor at three representative stages (WRL~3, WRL~5, WRL~7) for all four pillars, and Table~\ref{tab:rubric-column} reproduces a complete four-anchor column ($s=0,1,2,3$) for pillar~P2 (Cyber-Physical Systems Fluency) at the same three stages, so that at least one pillar is shown as a fully anchored ranked scale.

\begin{table}[H]
\centering
\caption{Illustrative behavioral anchors at three WRL stages. Each cell shows the anchor at score $s=2$ (``adequate/consistent'') as a representative point on the four-anchor scale; $s=0,1,3$ anchors bracket this value in the full rubric.}
\label{tab:rubric-selected}
\small
\begin{tabular}{@{}p{0.06\linewidth}p{0.20\linewidth}p{0.20\linewidth}p{0.20\linewidth}p{0.22\linewidth}@{}}
\toprule
\textbf{Stage} & \textbf{P1 Digital \& AI} & \textbf{P2 CPS Fluency} & \textbf{P3 Human--Machine} & \textbf{P4 Data-Driven Decision} \\
\midrule
WRL~3 & Reads a labeled ML dashboard and identifies one out-of-range feature. & Locates PLC I/O modules and traces a sensor signal to an HMI tag. & Identifies cobot stop categories and safety-rated zones from documentation. & Interprets an SPC chart and marks an out-of-control point with justification. \\
WRL~5 & Trains and evaluates an ML model on lab data with a documented train/val split and reports a defensible metric. & Integrates PLC~+~cobot~+~vision in a lab cell, publishes tags over OPC-UA, and documents the topology. & Runs a task-allocation exercise between operator and cobot with an ISO/TS~15066-aligned risk check. & Executes a DOE on a lab process, reports the analysis of variance (ANOVA), and recommends a setpoint. \\
WRL~7 & Deploys a monitored ML service against a production data stream with a rollback and drift plan. & Diagnoses and resolves a live OT/IT integration incident on the production network. & Leads a mixed human--cobot cell shift and documents an operator-facing standard-operating-procedure (SOP) change. & Runs a chartered continuous-improvement project with a measured KPI improvement. \\
\bottomrule
\end{tabular}
\end{table}

\begin{table}[H]
\centering
\caption{A complete four-anchor column for pillar~P2 (Cyber-Physical Systems Fluency) at three representative stages, illustrating the full ranked rubric ($s=0,1,2,3$) for one competency dimension. The full $9\times4\times4$ set of anchors is provided as Supplementary Material.}
\label{tab:rubric-column}
\small
\begin{tabular}{@{}p{0.05\linewidth}p{0.20\linewidth}p{0.22\linewidth}p{0.22\linewidth}p{0.22\linewidth}@{}}
\toprule
\textbf{Stage} & \textbf{$s=0$ (not observed)} & \textbf{$s=1$ (emerging)} & \textbf{$s=2$ (adequate)} & \textbf{$s=3$ (high standard)} \\
\midrule
WRL~3 & Cannot locate the PLC or identify its I/O. & Locates the PLC but mislabels I/O or cannot trace a signal. & Locates PLC I/O modules and traces a sensor signal to an HMI tag. & Traces multiple signals, explains scan-cycle timing, and flags a wiring fault unprompted. \\
WRL~5 & Cannot integrate the subsystems or configure communications. & Wires the cell but the OPC-UA tags are incomplete or undocumented. & Integrates PLC~+~cobot~+~vision in a lab cell, publishes tags over OPC-UA, and documents the topology. & Delivers a robust, fault-tolerant integration with alarm handling and a validated tag database. \\
WRL~7 & Cannot diagnose the live integration incident. & Identifies the incident but the fix is partial or destabilizes the line. & Diagnoses and resolves a live OT/IT integration incident on the production network. & Resolves the incident, implements a monitored preventive control, and documents the root cause for reuse. \\
\bottomrule
\end{tabular}
\end{table}

\end{document}